\documentclass{ar-1col}
\usepackage[numbers]{natbib}
\usepackage{url}
\usepackage{amsmath}
\usepackage{amssymb}

\jname{Xxxx. Xxx. Xxx. Xxx.}
\jvol{AA}
\jyear{YYYY}
\doi{10.1146/((please add article doi))}

\def\lsim{\lower.5ex\hbox{$\; \buildrel < \over \sim \;$}}

\def\gsim{\lower.5ex\hbox{$\; \buildrel > \over \sim \;$}}

\begin{document}

% Page header
\markboth{Tatischeff, Gabici}{Light element production by cosmic rays}

% Title
%\title{Particle acceleration by supernova remnants and production of light nuclei}
\title{Particle acceleration by supernova shocks and spallogenic nucleosynthesis of light elements}

%Authors, affiliations address.
\author{Vincent Tatischeff$^1$ and Stefano Gabici$^2$
\affil{$^1$Centre de Sciences Nucl\'eaire et de Sciences de la Mati\`ere (CSNSM), CNRS/IN2P3, Univ. Paris-Sud, Universit\'e Paris-Saclay, F-91405 Orsay Campus, France; email: vincent.tatischeff@csnsm.in2p3.fr}
\affil{$^2$APC Laboratory, Univ. Paris Diderot, CNRS/IN2P3, CEA/Irfu, Obs. de Paris, Sorbonne Paris Cit\'e, 75013 Paris, France; email: gabici@apc.in2p3.fr}
}

%Abstract
\begin{abstract}
In this review, we first reassess the supernova remnant paradigm for the origin of galactic cosmic rays in the light of recent cosmic-ray data acquired by the \textit{Voyager 1} spacecraft. We then describe the theory of light element nucleosynthesis by nuclear interaction of cosmic rays with the interstellar medium and outline the problem of explaining the measured Be abundances in old halo stars of low metallicity with the standard model for the galactic cosmic ray origin. We then discuss the various cosmic ray models proposed in the literature to account for the measured evolution of the light elements in the Milky Way, and point out the difficulties that they all encounter. 
Amongst all possibilities, it seems to us that the superbubble model provides the most satisfactory explanation for these observations.
%Abstract text, approximately 150 words. 
\end{abstract}

%Keywords, etc.
\maketitle

%Table of Contents
\tableofcontents

% Heading 1
\section{INTRODUCTION}
Galactic cosmic rays (GCRs) are widely believed to be accelerated at shock waves generated by supernova (SN) explosions in the interstellar medium (ISM)%, at least up to the ``knee'' energy of the CR spectrum at $\sim 10^{15}$~eV and possibly up to the ``ankle'' at $\sim 10^{15}$~eV 
\cite{blasireview,lukerecentreview}. In the last few decades, X- and gamma-ray observations of various CR candidate sources have brought valuable information on the particle acceleration mechanism at work in SN remnants (SNRs) \cite{hel12,fermipion,latsnrcat}. Observations of the diffuse Galactic emission with the \textit{Fermi} Large Area Telescope (LAT) have also shed new light on the CR origin and the propagation of these particles in the ISM \cite{latcocoon,latdiffuse}. However, these observations put constraints only on CRs of kinetic energies above a few hundred MeV/nucleon: the flux and composition of lower energy CRs in the ISM remain poorly known. 

Low-energy ($\lesssim$ GeV) CRs are thought to be major actors in the Galactic ecosystem. They are the primary source of ionization and heating of dense molecular clouds, and thereby play a pivotal role in driving interstellar chemistry and in the process of star formation in these regions. They may also excite magnetohydrodynamical turbulence in the ISM, and provide a critical pressure support to launch the Galactic wind into the halo \cite{gre15}. 
Here, we focus on the spallogenic nucleosynthesis of the light elements lithium, beryllium, and boron (LiBeB), induced by the nuclear interactions of sub-GeV GCRs with nuclei of the ISM \cite{ree70,men71}.
This process is important to explain the observed evolution of the LiBeB isotopes throughout the lifetime of the Galaxy \cite{van90,pra93a}. 
Remarkably, crucial information about the sources of GCRs can be obtained when predictions from this scenario are confronted with observations of light elements abundances found at the surface of old halo stars \cite{van00,ram00,pra12b}. 

The low-energy region of the interstellar CR spectrum cannot be directly measured near Earth because of the solar modulation effect. However, since August 2012, the \textit{Voyager 1} spacecraft has reached a region close to, or beyond the heliopause, and it is now measuring a stable flux of particles most likely corresponding to the local interstellar spectrum of CRs \cite{sto13}. The \textit{Voyager 1} scientific team has thus recently provided very valuable measurements of the local energy spectra of GCR nuclei down to 3~MeV/nucleon \cite{cum16}. When combined with the precision measurements of CRs of larger energies performed by the Alpha Magnetic Spectrometer (AMS-02) on the International Space Station \cite{ams1,ams2}, these data provide an unprecedented view of the local interstellar spectra of GCR nuclei over six orders of magnitude in energy. 
In addition to that, infrared and radio molecular lines have been detected in the spectra of a number of molecular clouds located within few kpc from the solar system. These lines are induced by the CR-ionization of interstellar molecular hydrogen and can be used to constrain the intensity of low-energy CRs \cite{ind12,neu17}. 

%Another crucial piece of information on the density of low-energy CRs in the ISM comes from the observations of infrared and radio spectral lines induced by the CR-ionization of interstellar molecules \cite{ind12,neu17}. 

In this review, we reassess the theory of light element production by GCR nucleosynthesis in the light of these new data on low-energy CRs. First, in Section 2 we discuss the SNR paradigm for the origin of GCRs \cite{hillas} and compare its predictions against the measurements of the local interstellar spectrum of GCRs as observed by \textit{Voyager 1} and AMS-02. We then perform in Section~3 a study of the spallogenic nucleosynthesis of light elements in the light of the measured CR spectra. In Section~4, we discuss the various CR models proposed in the literature to account for the measured evolution of the light elements in the Galaxy. We conclude in Section~5 with a brief summary. 

%Heading 1
\section{THE SUPERNOVA REMNANT PARADIGM FOR THE ORIGIN OF GALACTIC COSMIC RAYS}
Any theory for the origin of GCRs must explain how and where these particles take their energy from, why their energy spectrum is very close to a power law over so many decades in particle energy, the remarkably high level of isotropy of their observed intensity, and their chemical composition.

There is a universal consensus on the fact that the bulk of CRs (dominated by mildly relativistic nuclei of energy around 1 GeV/nucleon) is of Galactic origin, and a very wide consensus on the fact that such nuclei are accelerated at SNR shocks \cite{blasireview,lukerecentreview}. The main reasons supporting a SNR origin of CRs are quite solid and can be stated as follows \cite{hillas}:
\begin{enumerate}
\item{the mechanical energy released during the explosions of galactic SNe suffices to explain the observed intensity of CRs, provided that a (very plausible) efficiency of $\approx 10$\% is assumed for the conversion of mechanical energy into CRs \cite{baadezwicky,terhaar,strongCRpower};}
\item{an acceleration mechanism, diffusive shock acceleration, is believed to operate at SNR shocks and to produce particle distribution functions which are power laws in momentum $f_0(p) \sim p^{-q}$ with $q \approx 4$. This corresponds to energy spectra which are also power laws $\sim E^{-s}$ with $s \approx 2$ ($s \approx 1.5$) for relativistic (non relativistic) particle energies \cite{luke83};}
\item{the observed relative abundances of primary to secondary CR nuclei (for example the B/C ratio) indicate that the confinement time of relativistic particles in the Galaxy decreases with particle energy (or momentum) as $\tau_c \sim p^{-\delta}$, with $\delta \approx$ 0.3...0.8 \cite{strongreview,mau10}. The expected equilibrium spectrum of the CRs injected by SNRs in the Galaxy is then $\sim f_0 \tau_c \sim p^{-(q+\delta)}$, which is quite close to the observed one $\sim p^{-4.7}$ ($\sim E^{-2.7}$).}
\end{enumerate}

{\it In situ} observations of the Earth bow shock provide a direct proof of the occurrence of diffusive shock acceleration \cite{ellisonbowshock}.
On the other hand, to know whether diffusive shock acceleration operates at SNR shocks or not, one has to rely on indirect observations. The radio \cite{ginzburgsyrovatskii}  and X-ray \cite{koyamaelectrons} synchrotron emission detected from a large number of remnants demonstrates that SNR shocks do accelerate electrons (and thus plausibly also protons and nuclei) with power law spectra extending from the GeV up to the multi-TeV energy domain. 

The signature of the acceleration of protons and nuclei at SNR shocks can be found in their gamma-ray spectrum, which results from the decay of neutral pions produced during inelastic interactions between accelerated nuclei and the interstellar gas. A characteristic feature of the $\pi^0$-decay emission is a very sharp rise of the gamma-ray spectral energy distribution in the sub-GeV energy domain, which makes it clearly distinguishable from the emission from competing leptonic mechanisms (inverse Compton scattering and Bremsstrahlung) \cite{felixbook}. Such a distinctive feature has been detected from a handful of SNRs observed by the Fermi satellite (see e.g. Fig.~2 in \cite{fermipion}). Moreover, several SNRs interacting with dense molecular clouds have been detected in gamma rays in both the GeV and TeV energy domain \cite{felixSNRs}. The presence of the molecular cloud provides a dense target for inelastic proton-proton interactions, which induces a strong enhancement of the hadronic gamma ray emission from interacting SNR. For this reason, the spatial correlation observed between the gamma-ray emission and the gas density distribution definitely points towards a hadronic origin of the emission \cite{mereview}.
Thus, based on gamma-ray observations one can conclude that SNRs certainly accelerate nuclei up to (at least) the TeV energy domain.

Unfortunately, it remains unclear whether the acceleration mechanism operating at SNRs can boost the energy of particles all the way up to the PeV domain and beyond. 
In fact, nuclei must be accelerated by galactic sources up to an energy of $\sim 3 \times 10^{18}$ eV, where the transition from Galactic to extragalactic CRs is believed to occur.
The capability of SNRs to accelerate galactic CRs up to these extreme particle energies remains a major open issue that might question the validity of the SNR paradigm \cite{etienne}.
The problem is both observational and theoretical: on one side, gamma-ray observations beyond the TeV energy domain are difficult to be carried out due to very limited statistics \cite{felixreview}, and on the other, predictions from the theory of diffusive shock acceleration struggle to reach the extremely large particle energies required to fit CR observations \cite{klara}. 

In the remainder of this Section we discuss in some detail the mechanism of diffusive shock acceleration (Sec.~\ref{sec:DSA}) and its predictions for the spectral shape (Sec.~\ref{sec:escape}) and chemical composition (Sec.~\ref{sec:chemical}) of the CRs released by SNRs in the ISM. In Sec.~\ref{sec:SB} we present a variation of the SNR paradigm, where CRs are accelerated inside superbubbles, i.e., large cavities inflated in the ISM around star clusters, due to the combined effect of stellar winds and SN explosions.
The role of the transport of energetic particles through the ISM in shaping the spectrum of CRs observed locally are finally discussed in Sec.~\ref{sec:propagation}.

\subsection{Cosmic rays from supernova remnants: diffusive shock acceleration theory.}
\label{sec:DSA}

The assumption at the basis of the diffusive shock acceleration mechanism is that energetic particles are scattered very effectively by magnetic disturbances on both sides of the shock. Magnetic disturbances can be taken to be at rest in the fluid frame, where the scattering is assumed to be elastic. The scattering makes then the distribution function of energetic particles almost isotropic in such frame, which is essential in order to adopt a diffusive description of the problem.
While the energy of particles increases by a quite small amount after a single interaction with a shock, their diffusive behavior makes it possible for some of them to cross repeatedly the shock, and therefore increase substantially their energy. As we will show in the following, it is very remarkable that the energy spectrum of particles accelerated at shocks is independent on any detail of the scattering process, and depends on the shock compression only. Moreover, for strong shocks the solution has an universal behavior, and the particle distribution function at the shock becomes $f_0(p) \propto p^{-4}$.%(or $F_0(E) \propto E^{-2}$), 
%where $p$ is the particle momentum \cite{luke83}.

The picture described above fails to provide a good working framework when the velocity of the accelerated particle $v$ is comparable to or smaller than the shock speed. At such low particle energies the main issue is to understand how thermal particles advected into the shock can gain a sufficient amount of energy to enter (or be injected into) the acceleration mechanism. A discussion of the injection problem goes beyond the scope of this review, and the interested reader is referred to \cite{damianoinjection} for further details.

\begin{marginnote}[]
\entry{Shock compression factor}{is the ratio between the gas density downstream and upstream of a shock. For a shock propagating in a monoatomic gas is related to the shock Mach number $\cal M$ as $r = 4 {\cal M}^2/({\cal M}^2 +3)$. The compression factor is equal to 4 for strong (large Mach number)  shocks, and smaller than that for weak ones. }
\end{marginnote}
Consider now an infinite and plane shock that propagates at a constant speed $u_s$. %from right to left along the $x$-axis. 
It is convenient to move to the rest frame of the shock, where the upstream fluid enters the shock with a constant velocity $u_1 = u_s$, and the downstream one is carried away at $u_2 = u_s/r$, where $r$ is the shock compression factor. 
The essence of the physics of diffusive shock acceleration can be easily grasped after recalling that the acceleration proceeds due to repeated cycles of a CR around the shock. It has been shown in \cite{bell} that after each cycle the momentum of a particle is increased by a small amount $p_{i+1} = p_{i} (1+\beta_{\rm acc})$ with $\beta_{\rm acc} = 2 (u_1-u_2)/3 v \ll 1$, and that the particle undergoing acceleration has a small probability $P_{\rm esc} = 4 u_2/v \ll 1$ per cycle to be advected downstream and leave the system. Therefore, the number of particles that performed at least $k$ cycles, and then reached a momentum larger than $p = p_0 (1+\beta_{\rm acc})^k$ is $4 \pi p^2 f_0(>p) \sim (1-P_{\rm esc})^k$. After eliminating $k$ and differentiating with respect to $p$ one gets the power law solution:
\begin{equation}
\label{eq:acceleration}
f_0(p) = \frac{q ~ Q_0}{4 \pi p_{\rm inj}^3 u_s} \left( \frac{p}{p_{\rm inj}} \right)^{-q} ~~ {\rm where} ~~ q = 3+\frac{P_{\rm esc}}{\beta_{\rm acc}} = \frac{3r}{r-1}
\end{equation}
%where $q = 3r/r-1$. 
In deriving Equation~\ref{eq:acceleration} it was assumed that particles of momentum $p_{\rm inj}$ are injected into the acceleration mechanism at a rate per unit shock surface equal to $Q_0$.
It is a truly remarkable fact that Equation~\ref{eq:acceleration} does not depend on the properties of CR transport at shocks (e.g. the particle diffusion coefficient), but on the shock compression only. Moreover, a universal solution $f_0(p) \propto p^{-4}$ is obtained in the strong shock limit ($r = 4$).

The solution given by Equation~\ref{eq:acceleration} has been obtained by treating CRs as test particles, i.e., no reaction of the accelerated CRs onto the shock structure was considered. In fact, this approximation is questionable, because at least two kinds of non-linearities may enter the problem. First of all, under many (realistic) circumstances, the pressure of CRs cannot be neglected, and this leads to a modification of the shock structure that tends to increase the compression of the gas beyond $r = 4$ \cite{malkov}. 
Second, the streaming of CRs ahead of the shock leads to a plasma instability, called streaming instability, that excites magnetic turbulence, enhances the scattering of CRs, and may even significantly amplify the magnetic field strength at shocks \cite{klarareview}.
If these non-linearities are taken into account, deviations from the solution reported in Equation \ref{eq:acceleration} are found.
However, according to state-of-the-art models, the combined effect of shock modification and magnetic field amplification leads to spectra of accelerated CRs which are still remarkably close to power laws. More specifically, spectral slopes predicted by non-linear models are found to be slightly steeper than the canonical value $q = 4$ for strong shocks \cite{damianosteep}. These theoretical studies were prompted by the detection of several SNRs in gamma rays. These data seem to indicate that CR are accelerated at these objects with spectra steeper than $p^{-4}$ \cite{felixSNRs}.

\subsubsection{Particle escape from SNR shocks and the spectrum of CRs released in the ISM}
\label{sec:escape}

A crucial aspect of diffusive shock acceleration which is still far from being fully understood deals with the escape of CRs upstream of shocks. The understanding of the escape mechanism is mandatory in order to estimate the spectrum of CRs which are released in the ISM by SNRs (e.g. \cite{bell13}, see also \cite{meescape} for a review).

Clearly, in an idealized picture where the shock is assumed to move at constant speed and to be plane and infinite, no escape upstream is possible, and all the CRs accelerated at the shock will be eventually advected into the far downstream region. 
However, this is not a good description of a SNR shock, which has a finite (spherical) size and is characterized by a decelerated rather than constant expansion velocity.
According to the picture that emerged from a number of theoretical investigations, highest energy CRs are released at a very early phase of the SNR evolution, and particles of lower and lower energy are gradually released as the SNR shock speed decreases \cite{meescape}. 
Even though this qualitative picture seems quite plausible, a fully self-consistent and quantitative description of this continuous escape of particles is still lacking. This is especially true for the late phases of evolution of SNRs, and it is indeed possible that some of the lowest energy (i.e. below a not very well defined energy $E_{\rm break}$) CRs accelerated at the SNR shock are released at the end of the evolution of the remnant only, when the shock becomes subsonic and dissolves in the ISM \cite{pz05}.
The difficulties encountered in developing quantitative models reside in the non-linearity of the problem: it is the turbulence generated by the stream of accelerated CRs ahead of the shock that confines them into SNRs \cite{klarareview}.

Here we present a toy model that, despite of the (deliberate) simplifying assumptions, hopefully grasps all the main physical ingredients of the problem.
The central question to be addressed is: how does the spectrum of CRs released by SNRs in the ISM $F_{\rm esc}(p)$ relates to the spectrum of particles accelerated at the shock $f_0(p)$?
The relationship between these two quantities is derived below, under the assumption that at any given time during the SNR lifetime, a fixed fraction $\eta$ of the shock kinetic energy $(1/2) \varrho u_s^3$ is converted into CRs ($\varrho$ is the density of the ISM). In this scenario, the normalization factor in Equation~\ref{eq:acceleration} can be written as $Q_0 \sim \eta u_s^3$, as long as the slope of $f_0(p)$ stays in the range $4 < q < 5$ (which encompasses virtually all of the relevant situations).
 
%In fact, when discussing particle confinement it is more convenient to express spectra as a function of rigidity $R$ rather than momentum $p$.

%\begin{marginnote}[]
%\entry{Rigidity}{is defined as the particle Larmor radius $R_L$ multiplied by the magnetic field strength: $R = pc/Ze$, where $Z$ is the particle charge in units of the elementary charge $e$.}
%\end{marginnote}

A dimensional argument can be used to estimate the maximum momentum of CRs that can be confined within a spherical shock of radius $r_s$ expanding at a velocity $u_s$: a particle can be confined if its diffusion length upstream of the shock $D_B/u_s$ is smaller than the characteristic length scale of the system $r_s$ \cite{pz05}. Here, $D_B \sim r_g v$ is the Bohm diffusion coefficient, appropriate to describe particle transport in highly turbulent environments. $r_g$ is the particle gyroradius and $v$ its velocity. This implies:
$(pv)_{\rm max} \sim Z e r_s B_s (u_s/c)$
where $B_s$ is the magnetic field strength upstream of the shock and $Z$ is the particle charge in units of the elementary one $e$.
In both the relativistic ($v/c \sim 1$) and non-relativistic ($v/c \ll 1$) regimes the quantity $pv$ is very close to the kinetic energy of the particle. We can then rewrite the confinement condition in terms of the energy {\it per nucleon} as: $E_{\rm max} \approx (Z/A) e r_s B_s (u_s/c)$, where $A$ is the atomic mass.

The evolution of SNRs in an homogeneous medium is self similar, with $r_s \propto t^{\alpha}$ and $u_s = \alpha r_s/t$.
During the Sedov (energy conserving) phase $\alpha$ is equal to 2/5, while it is smaller than that during later phases, where radiative energy losses can no longer be neglected. In most scenarios for magnetic field amplification at shocks, a given fraction of the shock ram pressure $\varrho u_s^2$ is converted into magnetic energy $B^2/8 \pi$ \cite{klarareview}. It follows then, that the maximum energy per nucleon that can be confined at a SNR shock depends on time as $E_{\rm max} \sim t^{\delta}$, where $\delta = 3 \alpha -2$.

Consider now the acceleration upward in momentum at the shock, which can be described by a flux of particles per unit shock surface equal to $\phi(p) = (4 \pi/3) p^3 f_0(p) (u_1-u_2)$. The quantity $\phi(p_{\rm max})$ represents then the flux of particles that, being accelerated beyond $p_{\rm max}$, can escape the SNR. The spectrum of CRs released in the ISM during the whole lifetime of a SNR can finally be computed as:
\begin{equation}
F_{\rm esc}(p) \sim \left[ \phi(p_{\rm max}) (4 \pi r_s^2) \frac{{\rm d}t}{{\rm d} p_{\rm max}} \right]_{p_{\rm max} = p} \sim \eta p^{-q} (u_s^2 r_s^3)
\label{eq:escape}
\end{equation}
where all the proportionality constants have been dropped after the second equality.
The equation above is valid as long as $p_{\rm max}$ scales with time as a power law (regardless of the actual value of the slope $\delta$). The term $(u_s^2 r_s^3)$ is conserved during the Sedov phase ($\alpha = 2/5$), being proportional to the total energy of the system. %, while it becomes a declining function of time at later times (when $\alpha < 2/5$). %This is true as long as the acceleration efficiency $\eta$ stays constant with time. 

Equation~\ref{eq:escape} has been obtained here in the framework of a test-particle theory, but it remains valid also if non-linear effects are taken into account \cite{pz05}. It implies that, during the Sedov (energy conserving) phase, the spectrum of particles released in the ISM has the same shape than that found at the SNR shock: $F_{\rm esc}(p) \sim f_0(p)$. 
On the contrary, after a time $t_{\rm rad}$ the SNR enters the radiative phase ($\alpha < 2/5$) and the following things happen: {\it i)} $(u_s^2 r_s^3)$ is no longer constant, but decreases with time (or with particle momentum $p = p_{max}(t)$), because a non negligible fraction of the total available energy leaves the system in form of photons; {\it ii)} for the same reason, the shock compression factor increases well above the reference value $r = 4$, and as a consequence the spectrum of accelerated particles becomes harder than the canonical $p^{-4}$; {\it iii)} the assumption of constant acceleration efficiency becomes very questionable at late times, and $\eta$ is expected to drop when radiative losses are important and the shock Mach number approaches unity; {\it iv)} CR particles trapped in the SNR during the radiative phase may suffer ionization and Coulomb energy losses, inducing a hardening in their spectrum at low energies.  
It follows that, even if the details of the escape mechanism remain elusive, at an energy per nucleon equal to $E_{\rm max}(t_{\rm rad})$ a break in the spectrum of escaping CRs might be expected, in the form of a hardening of the spectrum towards lower momenta. 
%Moreover, the assumption of constant acceleration efficiency $\eta$ becomes very questionable at late times, when the shock Mach number approaches unity, and the hardening at low particle momenta in the spectrum of the escaping particles might even become more pronounced, due to a decline of $\eta$ with time. 

In the absence of both observational constraints and of a quantitative theory for escape, it is very difficult to predict at which particle momentum one should expect a deviation from $F_{\rm esc}(p) \sim f_0(p)$, as well as the amount of spectral suppression towards low energies.
Some theoretical studies seem to suggest that this might happen in the trans-relativitic or mildly relativistic ($\approx$ GeV domain) regime \cite{pz05}.
As we will show in the following, a low energy break in the spectrum of energetic particles injected in the ISM is indeed required in order to fit the observed local spectrum of CRs.
The qualitative results presented in this Section should be then considered as a plausibility argument in favor of the presence of such spectral break. 

\subsubsection{Chemical composition of primary cosmic rays at injection}
\label{sec:chemical}

\begin{figure}[t]
\includegraphics[width=0.8\textwidth]{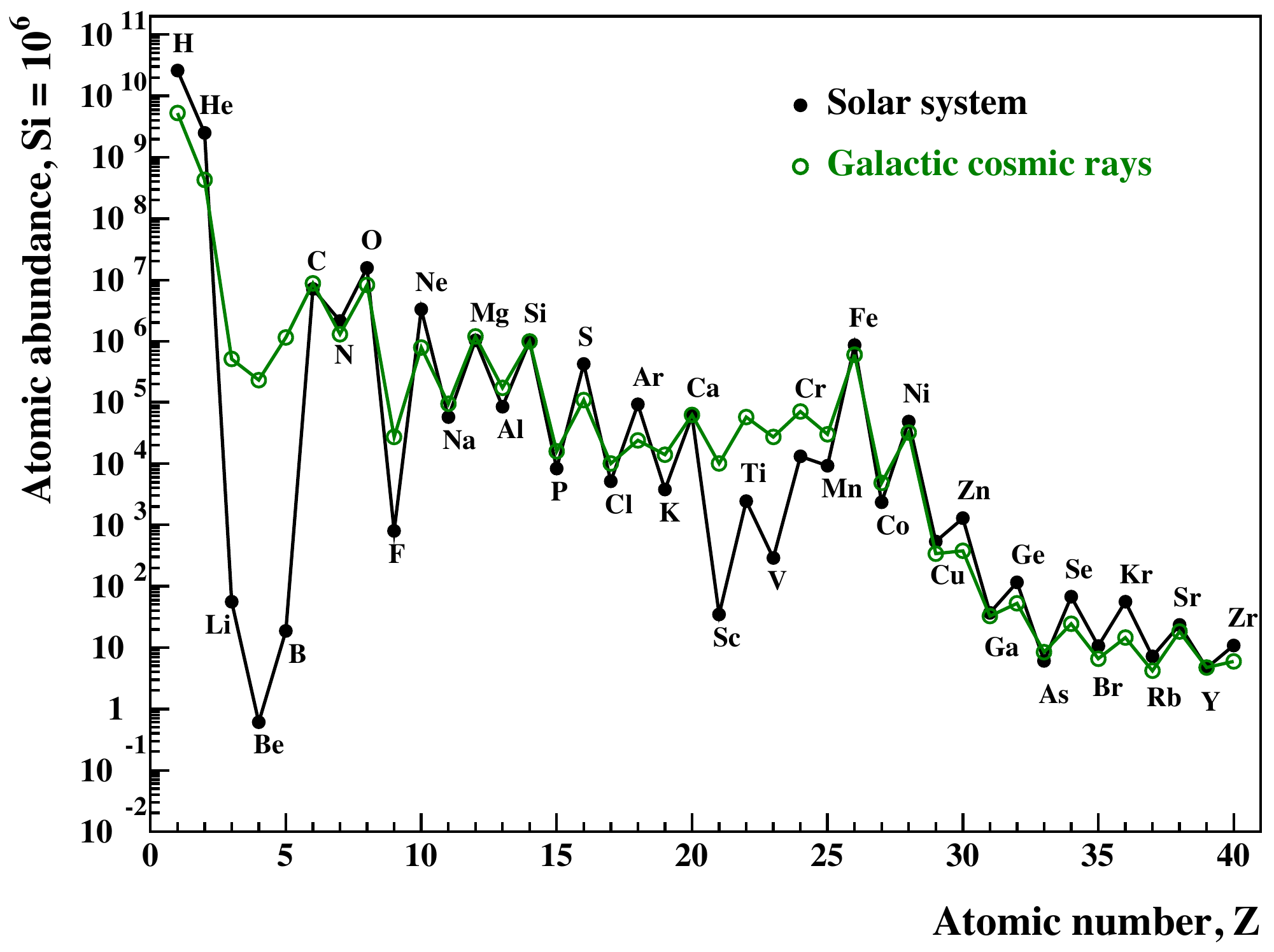}
\caption{Abundances of elements as function of atomic number up to $Z=40$ (Zr) normalized to $10^6$ Si atoms. The solar system abundances (black symbols) are taken from Table~10 in Ref.~\cite{lod09}. The GCR abundances (green symbols) are from \textit{Voyager 1} measurements \cite[][Table 3]{cum16} up to Fe ($Z=26$), from measurements with the Trans-Iron Galactic Element Recorder (TIGER) balloon-borne instrument \cite{rau09} for Co ($Z=27$) and Cu ($Z=29$), and from SuperTIGER observations \cite{mur16} for the other elements.}
\label{fig:abundances}
\end{figure}

The observed elemental abundances in GCRs are compared with the solar system abundances in \textbf{Figure~\ref{fig:abundances}}. Overall, the two distributions look remarkably similar for most elements, with some important exceptions.
In particular, light (Li, Be, B, the focus of this review) and sub-iron (Sc, Ti, V, Cr, Mn) elements are overabundant by many orders of magnitude in GCRs.
%, and their depletion in solar system material is due to the fact that they are essentially not produced in stellar nucleosynthesis.
%In particular, CRs are overabundant by orders of magnitude with respect to solar system abundances for light (Li, Be, B, the focus of this review) and sub-iron nuclei. 
This striking difference is interpreted as the result of nuclear spallation reactions that occur during the propagation of CRs in the ISM, resulting in the breakup of relatively abundant heavy nuclei into lighter ones \cite{ginzburgsyrovatskii}.

A less pronounced but still very evident difference is the enhanced abundance in the cosmic radiation of elements with $Z > 2$ with respect to H and He. Such enhancement is not related to nuclear processes, but is rather regulated by atomic parameters such as the mass-to-charge ratio. In the same line, refractory elements are found to be relatively more abundant than volatile ones in GCRs
\cite{wiedenbeck,rau09,mur16}, which can be explained in a scenario where dust grains, being characterized by a very large mass-to-charge ratio, are accelerated very effectively at shocks \cite{mey97,ell97}. During the acceleration, grains attain velocities large enough to be eroded by sputtering. The sputtered particles would then be refractory elements, that will have the same velocity of the parent grain. Such a velocity is much larger than the shock speed, and this guarantees the injection of refractory elements ejected by grains into the acceleration process, independently on their mass-to-charge ratio \cite{mey97,ell97}.

\begin{marginnote}[]
\entry{Volatility}{is the tendency of an element to be found in its gaseous state, rather than condensed into dust grains. Elements with low (high) condensation temperature are called volatiles (refractory).}
\end{marginnote}

\begin{marginnote}[]
\entry{Rigidity}{regulates the motion of particles in a magnetic field $B$, and is defined as $R = r_g B = \frac{p c}{Z e}$ where $r_g$ is the particle gyration radius.}
\end{marginnote}

Among the GCR volatile elements, the heavier ones are found to be relatively more abundant than lighter ones, while such a trend is not observed (or is very much weaker) among refractory elements \cite{rau09,mur16}. From theory, it is difficult to see how the atomic mass $A$ alone could be the physical parameter regulating the acceleration efficiency of volatile elements. A much more plausible physical parameter would be the rigidity, which is proportional to the mass-to-charge ratio $A/Z$ of ions, and governs the behavior of particles in magnetized environments such as shocks. %plays a central role in the diffusive shock acceleration mechanism.
Indeed, such a rigidity dependent enhancement is predicted by state-of-the art simulations of diffusive shock acceleration, with a scaling in the sub-relativistic particle energy domain equal to \cite{cap17}: 
\begin{equation}
\label{eq:enhancement}
C_i \equiv \frac{f_i(E/Z_i^{\rm ISM})}{\chi_i f_p(E)} \sim (A_i/Z_i^{\rm ISM})^2~.
\end{equation}
Here, $f_i$ and $f_p$ are the CR particle distribution functions at the shock for elements of specie $i$ of atomic mass $A_i$ and for protons, respectively, $\chi_i$ is the ISM abundance of element $i$ with respect to hydrogen, $E$ is the particle energy, and $Z_i^{\rm ISM}$ represents the typical charge (ionization state) of ions $i$ in the ISM at the moment on their injection into the acceleration process.
The value of $Z_i^{\rm ISM}$ is in general different from $Z_i$ and typically is equal to 1 in the warm phase of the ISM, and larger than that in the hot one. 
The enhancement factor can be parametrized as $\sim A_i^{\alpha}$, with $\alpha = 2$ for singly ionized elements, and $\alpha < 2$ for higher levels of ionization, under the assumption that the charge state $Z_i^{\rm ISM}$  scales with some power of the atomic mass $A_i$.
Ions of charge $Z_i^{\rm ISM}$ (generally smaller than $Z_i$) are injected from the ISM into the acceleration process, and they are then stripped of their electrons as the acceleration proceeds to large energies \cite{morlino}.

Here and in the following we will focus mainly on H-He and CNO CRs, since they dominate the production of the light nuclei Li, Be, and B in direct and inverse spallation reactions, respectively. All these elements are highly volatile, and their abundance in GCRs should then be enhanced according to their mass-to-charge ratio as described by Equation~\ref{eq:enhancement}. We note that the observed overabundance of the $^{22}$Ne isotope in GCRs suggests that a significant fraction of these particles comes from the acceleration of Wolf-Rayet wind material enriched in He-burning products, mainly $^{12}$C, $^{16}$O and $^{22}$Ne \cite{mey97}. This has no impact on the LiBeB calculations, but it will be further discussed, however, in Section~\ref{sec:gcrmodelwind}.
%The enhancement of CRs of atomic mass $A$ can be conveniently expressed as:
%\begin{equation}
%\label{eq:enhancementp}
%f_A(p) \sim A~ \chi_A~ p^{-q}
%\end{equation}
%where we have assumed an interstellar medium constituted by singly ionised atoms ($Z_{\rm ISM} = 1$) and we used particle momentum rather than energy as a variable, to make the Equation~\ref{eq:enhancementp} valid at both relativistic and non-relativistic particle energies.

%Following the assumptions of Ref.~\cite{cap11} for the particle injection into the diffusive shock acceleration process, the CR source spectrum as a function of kinetic energy per nucleon (expressed in number of particles cm$^{-3}$ s$^{-1}$ (MeV/nucleon)$^{-1}$) can be written as
%\begin{equation}
%\dot{Q}_i(E)  = {C K_i Z_i^{s-5} \over \beta} p^{-s}~,
%\label{eq:q}
%\end{equation}
%where $C$ is a constant independent of the accelerated particle species, $K_i$ is the source abundance of the fast ions of type $i$ relative to protons, $Z_i$ is the nuclear charge of these particles, $\beta=v/c$ is the particle velocity relative to that of light, and $p$ the particle momentum per nucleon. 

\subsection{Cosmic ray acceleration in superbubbles}
\label{sec:SB}

In describing the SNR paradigm for the origin of CRs we implicitly assumed that the acceleration of particles takes place at the shocks of {\it isolated} SNRs, propagating in an {\it average} ISM of density $\sim 1$~cm$^{-3}$. While this is a very plausible assumption for thermonuclear (Type Ia) SNe, it is most likely not a good one for core-collapse SNe, whose progenitors are very often found in OB stellar associations. Massive O and B stars form in groups (associations) as the result of the collapse of giant molecular clouds. Estimates of the fraction of O and B stars found in such associations range from $\sim 60$\% \cite{garmany} to $\sim$95\% \cite{hig98}. This, together with the fact that core-collapse SNe constitute a vast majority of all of the galactic SN explosions, implies that a large fraction of Galactic SNRs originated in OB association, and therefore that the explosion of SNe in the Galaxy is likely to be quite strongly correlated in both time and space. 

In this scenario, the shocks of SNRs originating from OB associations propagate in the large ($\gg 10$ pc), hot ($> 10^6$ K), diluted ($< 10^{-2}$ cm$^{-3}$), and highly turbulent cavities inflated by the combined effect of stellar winds and SN explosions of the OB stars that belong to the association \cite{kaf81}. Such cavities, called superbubbles, are surrounded by dense shells of swept-up ISM, and their interior consists of a mixture of SN ejecta and stellar wind material from OB stars, with the addition of some standard ISM that survived inside the cavity in the from of dense clumps, or that evaporated off the dense shell \cite{mac88}. The chemical enriched interior may play an important role in the spallogenic nucleosynthesis of LiBeB, as discussed in Section \ref{sec:SBLiBeB}, and their large size and highly turbulent interior might confine and accelerate nuclei up to the energies of the transition to extragalactic CRs \cite{etienne}.

Particles can be accelerated at individual SNRs inside superbubbles via diffusive shock acceleration. In this case the acceleration would be concentrated in the relative small fraction of the volume of the cavity surrounding the most recent SNe \cite{hig98}.
On the other hand, the collective effects of stellar winds and SN explosions would make the cavity an extremely turbulent medium, characterized by the presence of multiple shocks, intermittence, large scale turbulent flows, and MHD waves. Particles can be accelerated due to the interaction with this complex ensemble of flows, and in this case the acceleration would be distributed more or less throughout the entire cavity \cite{byk90,byk92,byk14}.

A rigorous description of the acceleration of particles in such a medium is not at all straightforward, due to the poorly constrained properties and very complex structure of the turbulent fluid. However, an analogy can be made with the acceleration of CRs at multiple shock waves spread over the entire volume of the cavity and characterized by a stochastic distribution of velocities \cite{par00,par04}. A characteristic particle energy $E_{\rm break}$ can be identified by equating the particle diffusion length ahead of a shock to the typical distance between shocks. Particles with energy $< E_{\rm break}$ have a very low probability to escape the shocks, because a particle advected away from a given shock would be caught and further accelerated by another one. A qualitative insight onto the particle spectrum resulting from this acceleration mechanism can be obtained from Equation~\ref{eq:acceleration}. After assuming that no escape of particle from the acceleration region is possible ($P_{\rm esc} = 0$) one obtains a very hard spectrum described by $f_0(p) \sim p^{-3}$ \cite{par00}. While the hardening of the spectrum below $E_{\rm break}$ is a quite solid prediction, the exact numerical value of $E_{\rm break}$ and of the spectral slope depend on the poorly constrained details of the acceleration mechanism.
However, models performed by independent groups suggests that $E_{\rm break}$ should fall in the trans-relativistic energy domain \cite{byk14,ferrand}.
For energies larger than $E_{\rm break}$ the acceleration mechanism changes significantly and is no longer due to the interaction with multiple shocks. In particular, detailed calculations showed that the spectrum steepens, resembling that expected from the acceleration of particles at a single shock: $f_0(p) \sim p^{-q}$ with $q \gsim 4$ \cite{byk14}. %$q \gtrsim 4$ (vt)
This finding suggests that, as found for SNRs, also in the superbubble scenario a low energy break might be present in the spectrum of CRs released in the ISM.

\subsection{Cosmic ray propagation in the interstellar medium: the leaky-box model.}
\label{sec:propagation}

Once released from their sources, CRs propagate in the turbulent and magnetized ISM and, depending on their energy, may eventually escape the Galaxy, lose their energy or be destroyed due to interactions with ISM matter.
The relative abundances of primary and secondary CRs (e.g. the B/C ratio) can be explained in terms of spallation reactions if the bulk of CRs ($\approx$ GeV particles) traverse on average a grammage $\Lambda_{\rm esc} \approx 10$~g/cm$^2$ before escaping the Galaxy. For relativistic particles and a typical mass density $\varrho_{\rm ISM} \sim 10^{-24}$~g~cm$^{-3}$, the residence time $\tau_{\rm esc} \sim 10$~Myr. It follows that the path length traveled by CRs before escape $L = \tau_{\rm esc} v \sim 3$~Mpc, which is way larger than the size of the Galaxy. 
This suggests that the confinement of CRs in the Galaxy is diffusive, and provides an explanation for the high level of isotropy observed in the arrival direction of CRs at Earth \cite{lukerecentreview}.

The diffusive behavior of CRs is due to their interactions with the turbulent interstellar magnetic field. A detailed analysis of this process goes beyond the scope of this review and in the following we present a simplified and stationary model where CRs are assumed to be confined within a given volume (a box) and have a constant probability per unit time to escape (or leak) out of it. This is called {\it leaky box model}, and is described by \cite{men71}:
\begin{marginnote}[]
\entry{Grammage}{is equivalent to a thickness and is generally expressed in g/cm$^2$. It is related to the residence time of CRs in the Galaxy, $\tau_{\rm esc}$, by $\Lambda_{\rm esc} = \tau_{\rm esc} v \varrho_{\rm ISM}$, where $v$ is the CR particle velocity and $\varrho_{\rm ISM}$ the average mass density of the ISM crossed by CRs before escaping.}
\end{marginnote}
\begin{equation}
\label{eq:leaky}
\frac{N_i(E)}{\tau_i^{\rm tot}(E)} + \frac{\partial}{\partial E} \left[ \left( \frac{{\rm d} E}{{\rm d} t} \right)_i N_i(E) \right] = \dot{Q}_i(E)~, 
\end{equation}
which can be solved to obtain the equilibrium distribution of CRs $N_i(E)$ of specie $i$ as a function of the energy per nucleon $E$. Particles of a given specie are assumed to be injected in the ISM at a rate $\dot{Q}_i(E)$ due either to the escape from astrophysical accelerators or as the result of the breakup of heavier species in spallation reactions. They are then subject to  both continuous (mainly ionisation and Coulomb energy losses at a rate $({\rm d} E/{\rm d} t)_i$) and catastrophic losses on a characteristic time scale $\tau_i^{\rm tot}$, which generally include the effects of escape from the Galaxy and nuclear destruction (as well as radioactive decay for radioisotopes): 
\begin{equation}
\frac{1}{\tau_i^{\rm tot}} = \frac{1}{\tau_i^{\rm ine}} + \frac{1}{\tau_i^{\rm esc}}
\cong \frac{1}{v(n_{\rm H}\sigma_{i{\rm H}}^{\rm ine} + n_{\rm He}\sigma_{i{\rm He}}^{\rm ine})} + 
\frac{\Lambda_{\rm esc}}{v(n_{\rm H}m_{\rm H}+n_{\rm He}m_{\rm He})}~.
\end{equation}
%with 
%\begin{equation}
%\tau_i^{\rm ine} \cong \frac{1}{v(n_{\rm H}\sigma_{i{\rm H}}^{\rm ine} + n_{\rm He}\sigma_{i{\rm He}}^{\rm ine})}~~~~~~
%\end{equation}
%{\rm and} ~~~~~~
%\begin{equation}
%\tau_i^{\rm esc} \cong \frac{\Lambda_{\rm esc}}{v(n_{\rm H}m_{\rm H}+n_{\rm He}m_{\rm He})}~.
%\end{equation}
Here, $n_j$ is the density of the ambient constituent $j$,  $\sigma_{i{\rm H}}^{\rm ine}$ and $\sigma_{i{\rm He}}^{\rm ine}$ are the total inelastic cross sections \cite[see][and references therein]{web90,tri99,mos02} for particle $i$ in H and He (by far the most abundant constituents in the ISM), and $m_{\rm H}$ and $m_{\rm He}$ are the H- and He-atom masses. 

%\subsubsection{The leaky box model}

Equation~\ref{eq:leaky} can easily be solved for stable, primary nuclei. If the contribution from spallation is neglected in the injection term, then $\dot{Q}_i(E)$ represents the spectrum of CRs injected in the ISM by astrophysical sources\footnote{This is a very accurate assumption for p, He, C, and O, while it is valid only in an approximate way for N \cite{wiedenbeck}.}.
The equilibrium solution for the CR energy spectrum in the ISM is then \cite{men71}:
\begin{equation}
N_i(E) = \frac{1}{({\rm d} E/{\rm d} t)_i(E)} \int_E^{\infty} dE' \dot{Q}_i(E') P_i(E',E)~,
\label{eq:n}
\end{equation}
where $P_i(E',E)$ is the survival probability of particles $i$ in the Galaxy as they slow down from $E'$ to $E$:
\begin{equation}
P_i(E',E) = \exp \bigg[- \int_E^{E'} \frac{dE''}{({\rm d} E/{\rm d} t)_i(E'') \tau_i^{\rm tot}(E'')}\bigg]~.
\label{eq:p}
\end{equation}

In the following, we compute the ionization and Coulomb energy loss rates ~\cite{schlibook} assuming that 20\% of the ISM density is in form of ionized H of temperature $10^4$~K.  

\subsubsection{A fit to Voyager data}

\begin{figure}[t]
\includegraphics[width=0.99\textwidth]{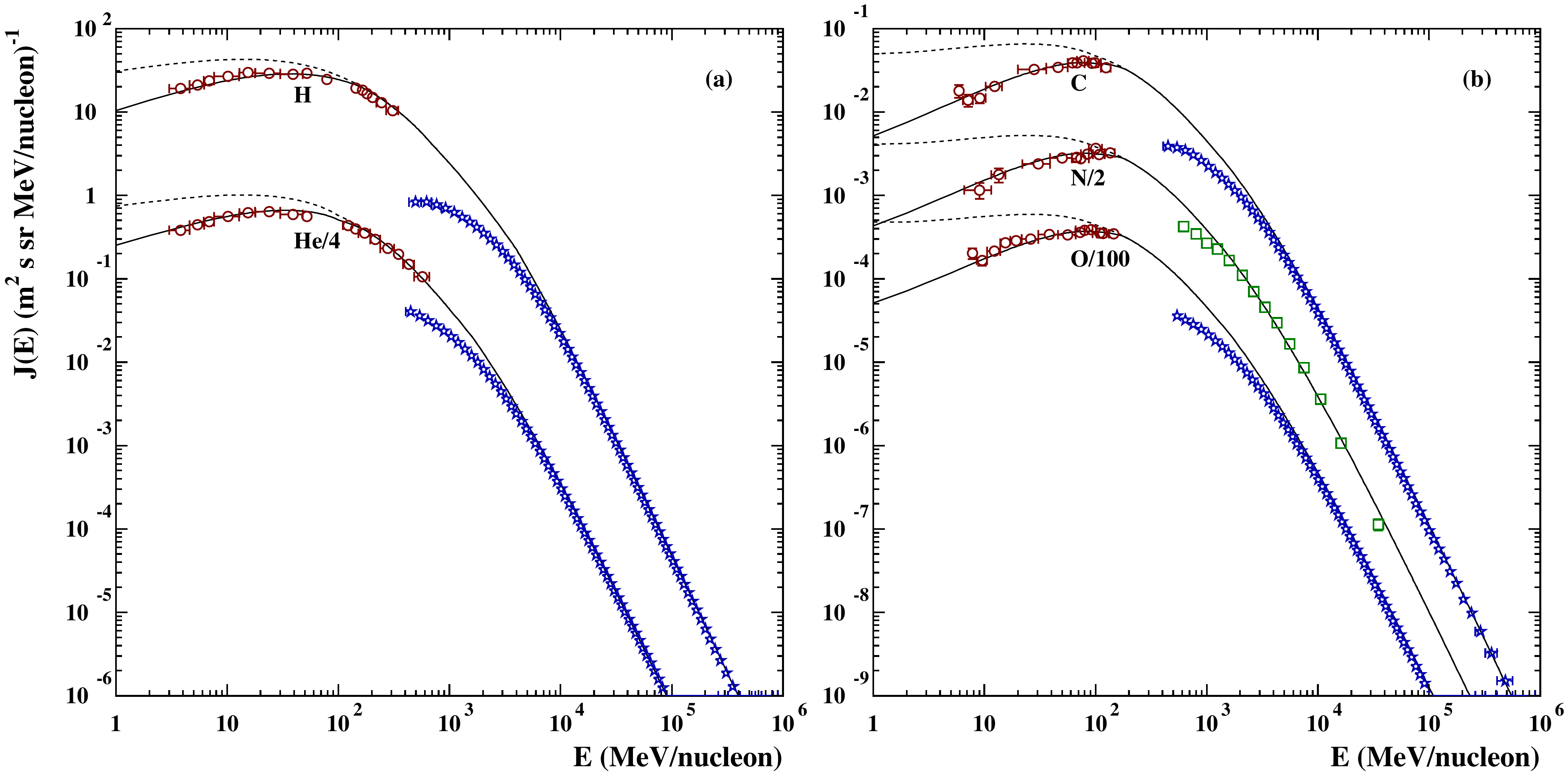}
\caption{Differential fluxes of GCR protons and $\alpha$-particles (panel a) and of CNO nuclei (panel b), measured by \textit{Voyager 1} (red circles) in the local ISM \cite{cum16}, AMS-02 \cite[blue stars;][]{ams1,ams2} and HEAO-3-C2 \cite[green squares;][]{eng90} near Earth. The solid curves show the best-fit propagated spectra from the leaky-box model assuming a broken power-law source spectrum: $E_{\rm break}=200$~MeV/nucleon, $q_{\rm h.e.}=4.3$, $q_{\rm l.e.}=3.75$ for H and He, and 3.0 for CNO (see text). The dashed lines show the results for an unbroken power-law source spectrum of slope $q=4.3$.}
\label{fig:gcrspectra}
\end{figure}

\textbf{Figure~\ref{fig:gcrspectra}} shows the local interstellar spectrum of CRs as observed by \textit{Voyager~1} \cite{cum16}, together with near Earth (i.e. subject to solar modulation) measurements of CRs performed by AMS-02 \cite{ams1,ams2} and HEAO-3-C2 \cite{eng90}. 
In order to interpret {\it Voyager~1} data, we rely on the earlier results of Jones et al. ~\cite{jon01},
%, which shows that the leaky box model is equivalent to the standard disk-halo diffusion model of CR transport, if the escape path length is related to the rigidity-dependent particle diffusion $D(R)$ as 
%\begin{equation}
%$\Lambda_{\rm esc} = {\mu \beta c H / 2 D(R)}~$.
%\label{eq:lambda}
%\end{equation}
%Here, $\mu \approx 2.4$~mg cm$^{-2}$ \cite{fer98} is the surface mass density of the Galactic disk and $H$ the CR halo boundary, i.e. the distance from the Galactic plane at which CRs freely exit from the Galaxy. 
where best fits to near Earth measurements of the B/C and sub-Fe/Fe GCR abundance ratios were obtained after adopting the following grammage:
\begin{equation}
\label{eq:grammage}
\left( \frac{\Lambda_{\rm esc}}{{\rm g/cm}^2} \right) = \left\{ \begin{array}{ll}
11.8 ~\beta  & {\rm~~for~} R <4.9 ~{\rm GV} \\
11.8 ~ \beta~ (R/4.9 ~{\rm GV})^{-0.54} & {\rm~~for~} R \ge 4.9 ~{\rm GV}~,
\end{array} \right.
\end{equation}
%with $\Lambda_0=11.8$~g~cm$^{-2}$, $R_0=4.9$~GV and $a=0.54$.
where $\beta = v/c$ is the CR velocity in units of the speed of light and $R$ is the particle rigidity.
Note that, as expected in most propagation scenarios, the scaling found at low rigidities ($\Lambda_{\rm esc} \propto \beta$) corresponds to a constant escape time $\tau_{\rm esc}$, i.e., to an escape of CRs from the Galaxy regulated either by advection or by a rigidity independent diffusion. This fact makes very plausible an extrapolation of Equation~\ref{eq:grammage} down to the rigidities probed by {\it Voyager~1}.
Having fixed $\Lambda_{\rm esc}$, Equation~\ref{eq:leaky} can be solved to fit data and eventually derive the injection spectrum of CRs in the ISM $\dot{Q}_i(E)$. 
%the only remaining free parameters of the leaky box model are the slope $s$ and abundances $K_i$ of the CR source spectrum (equation~\ref{}).

As a first attempt, we assume a pure power law in momentum per nucleon for the CR injection spectrum:
\begin{equation}
\dot{Q}_i(E) = 4\pi p^2 f_i(p) \frac{{\rm d}p}{{\rm d}E}= K_i \beta^{-1} \left(\frac{p}{A_i} \right)^{2-q}
\label{eq:q}
\end{equation}
and we obtain a good fit to high energy data (dotted lines in \textbf{Figure~\ref{fig:gcrspectra}}) for $q = 4.3$ and for $K_i/K_H = 0.095, 0.0039, 0.00072, 0.0046$ for $i = $ He, C, N, and O, respectively. 
In order to fit also \textit{Voyager~1} data (solid lines in \textbf{Figure~\ref{fig:gcrspectra}}) we introduce a break in the injection spectrum of all species at a fixed energy per nucleon: $E_{\rm break} = 200$ MeV/nucleon. We keep the slope of the spectrum above the break equal to $q_{\rm h.e.} = 4.3$, while below $E_{\rm break}$ we found $q_{\rm l.e.} = 3.75$ for H and He and $q_{\rm l.e.} = 3.0$ for CNO nuclei (the pedices h.e. and l.e. stand for high energies and low energies, respectively). A low energy break is indeed expected in most scenarios of acceleration of CRs (see Sections~\ref{sec:escape} and \ref{sec:SB}), but the reason of the difference of $q_{\rm l.e.}$ for different species remains unclear and should be further investigated.
This difference also makes it difficult to perform a direct comparison between our results and the prediction of diffusive shock acceleration theory for the CR chemical composition (Section \ref{sec:chemical}).
The CR injection spectrum $\dot{Q}_i(E)$ obtained from the fit to data will be used in the following to make predictions about the spallogenic nucleosynthesis of light elements.

\section{NONTHERMAL SYNTHESIS OF THE LIGHT ELEMENTS Li, Be, AND B} 
\subsection{Observations}

The abundance curve of the elements in the solar system presents a remarkable gap between He and the CNO elements (\textbf{Figure~\ref{fig:abundances}}), which suggests that the rare elements Li, Be, and B (LiBeB) are produced  in the Universe by a specific process of nucleosynthesis. Already in the B$^2$FH landmark paper on stellar nucleosynthesis \cite{B2FH}, the authors pointed out that the LiBeB isotopes are rapidly destroyed in stellar interiors by thermonuclear reactions with protons, and suggested that these isotopes are synthesized by a unknown ``x-process'' in a low-temperature, low-density environment. 

Since then, it was found that primordial nucleosynthesis (or Big Bang Nucleosynthesis, BBN) synthesized a significant amount of $^7$Li \cite[e.g.,][]{ree94}, as evidenced by the Spite plateau \cite{spi82} of Li abundances in low metallicity halo stars of the Galaxy\footnote{Standard BBN calculations using the cosmological parameters determined by the \textit{Planck} satellite mission predict a Li primordial abundance about three times larger than the abundance deduced from spectroscopic observations of metal-poor halo stars, and the origin of this discrepancy is not yet understood \cite[see, e.g.,][and references therein]{coc17}.}. But the predictions of standard BBN for the primordial abundances of the other LiBeB isotopes ($^6$Li, $^9$Be, $^{10}$B and $^{11}$B) are at least three orders of magnitude below the abundances measured in metal-poor stars \cite{coc14}.

The first hint that Li, Be, and B are significantly produced by nuclear interaction of GCRs with the ISM comes from the measured abundances of these species in the GCRs, which, relative to, e.g., Si, are higher than the solar system abundances by factors ranging from about 10$^4$ for Li to 10$^6$ for Be (see \textbf{Figure~\ref{fig:abundances}}). However, the first quantitative study of this idea had to wait for accelerator measurements of the cross sections for the nuclear spallation reactions leading to the formation of LiBeB from proton bombardment of $^{12}$C, $^{14}$N, and $^{16}$O, which were performed in Orsay in the mid-1960s \cite{ber67}. In 1970, Reeves et al. \cite{ree70} showed that the pre-solar abundances of the light elements as measured in meteorites can be approximately reproduced by assuming that they are synthesized by fragmentation of interstellar (or GCR) CNO by GCR (or interstellar) protons and $\alpha$-particles, plus a contribution from $\alpha+\alpha$ reactions for Li production, and further assuming that the GCR flux and CNO abundances have not changed throughout the history of the Galaxy. Quantitatively, Reeves et al. \cite{ree70} found that a GCR proton flux of $F_p(E>30~{\rm MeV})\sim12$~cm$^{-2}$~s$^{-1}$ averaged over $\sim$12~Gyr would be needed to reproduce the pre-solar abundance of Li (\textit{sic}) and Be. Remarkably, the proton flux measured by \textit{Voyager 1} in the local ISM (\textbf{Figure~\ref{fig:gcrspectra}}) is $F_p(E>30~{\rm MeV})=15$~cm$^{-2}$~s$^{-1}$. 

\begin{marginnote}[]
\entry{Metallicity}{is the fraction of mass of a star or any astrophysical object that is not in H and He. The protosolar metallicity is $Z_\odot=0.0153$ \cite{lod09}. The metallicity is often estimated from proxies such as the O or Fe abundances and the following notation is often used: [Fe/H] = $\log[{\rm (Fe/H)/(Fe/H)}_\odot]$, where (Fe/H)/(Fe/H)$_\odot$ is the Fe number abundance relative to its solar value.}
\end{marginnote}

More detailed calculations of LiBeB production, taking into account the diffusive transport of GCRs in the Galaxy, were published one year later by Meneguzzi et al. \cite{men71}. These authors found that the GCR productions of $^7$Li and $^{11}$B are not enough to account for the meteoritic data, and they proposed two additional sources for these isotopes: (i) stellar nucleosynthesis in giant stars and (ii) spallogenic nucleosynthesis by an \textit{ad hoc} low-energy CR component, called ``carrot'', not detectable near Earth due to the solar modulation. Such a supplementary flux of low-energy CRs could significantly enhance the production of $^7$Li by $\alpha+\alpha$ reactions and that of $^{11}$B by the $^{14}$N($p$,$x$)$^{11}$B reaction (see the cross sections for these reactions in panels (g) and (b) of \textbf{Figure~\ref{fig:siglibeb}}, respectively). 

Since the seminal paper by Meneguzzi et al. \cite{men71}, three possible stellar sources of $^7$Li were identified \cite[see][and references therein]{pra12b}: asymptotic giant branch stars (from $^3$He+$^4$He reactions in the bottom of the convective stellar envelope), classical novae (from explosive H-burning), and core-collapse SNe (from $\nu$-induced nucleosynthesis). Thus, $^7$Li is the only isotope in nature produced by either BBN, stellar nucleosynthesis, or GCR interaction. A significant stellar production of $^{11}$B is thought to arise from  $\nu$ spallation of $^{12}$C in core-collapse SNe \citep{woo90,nak10}, which could account for the missing source of $^{11}$B identified by  Meneguzzi et al. \cite{men71}. At present, only $^6$Li, $^9$Be and $^{10}$B are thought to be pure products of GCR nucleosynthesis\footnote{Asplund et al. \cite{asp06} reported observations of high $^6$Li abundances in metal-poor halo stars unexplainable by GCR nucleosynthesis, which would require an additional source also for this isotope. Several production scenarios were proposed in the literature: (i) non-standard BBN \cite[e.g.,][and references therein]{jed09}, (ii) pre-galactic nucleosynthesis during structure formation \cite[e.g.,][]{rol06} or (iii) \textit{in situ} production by stellar flares \cite{tat07}. But these high $^6$Li abundances were not confirmed by subsequent observations \cite{lin13}.}.

\begin{figure}[t]
\includegraphics[width=0.95\textwidth]{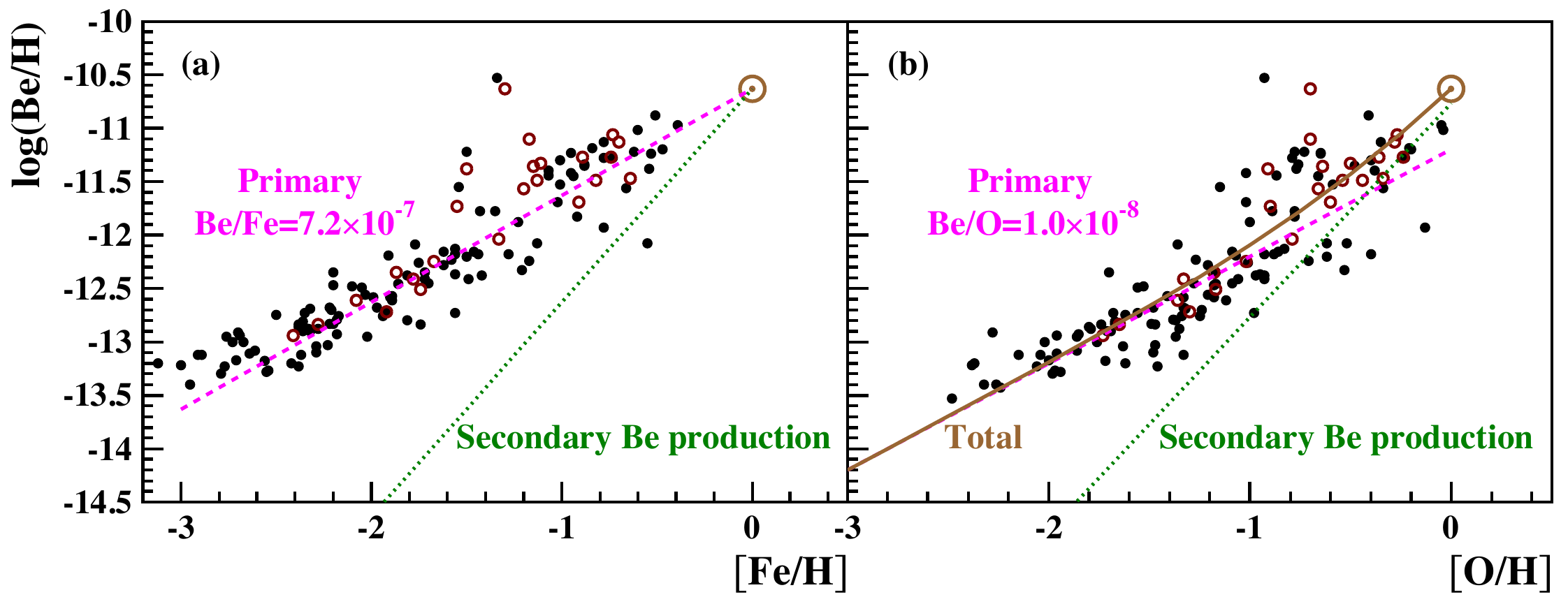}
\caption{Observations of Be abundances vs. [Fe/H] (panel a) and [O/H] (panel b). Data are from Refs.~\cite[][black dots]{boe11} and \cite[][red circles]{tan09}. The dashed (dotted) lines of slope one (two) represent a primary (secondary) Be production. In panel (b), the solid line shows a fit to the data with a secondary Be production component plus a primary component of constant $\rm{Be/O} = 1.0 \times 10^{-8}$. The Be vs. Fe data (panel a) are consistent with a constant $\rm{Be/Fe} = 7.2 \times 10^{-7}$.}
\label{fig:be9data}
\end{figure}

Studies of the origin of LiBeB took an unexpected turn in the 1990s, when further observations of metal-poor halo stars \cite[e.g.][]{boe93,dun97} revealed that Be and B abundances increase \textit{linearly} with [Fe/H]. This is illustrated for the Be evolution in \textbf{Figure~\ref{fig:be9data}a} with the data of Refs.~\cite{boe11,tan09}; measurements by other groups \cite[e.g.,][]{smi09,pri10} show a similar trend \cite[see Figure 1 in Ref.][]{pra12b}. The observed metallicity dependence of Be was unexpected, because this element was thought to be synthesized by spallation of increasingly abundant CNO nuclei in both the ISM and the GCRs, which is a \textit{secondary} production process leading to a \textit{quadratic} dependence of the nucleosynthesis product with metallicity \cite[see][]{van90,van00}. 

\begin{marginnote}[]
\entry{Primary and secondary element production}{in galactic chemical evolution theory, \textit{primary} elements have production rates independent of metallicity, because they are synthesized directly from H and He, while the production rates of \textit{secondary} elements depend on the initial metallicity of their progenitors \cite{pag}. Not to be confused with the \textit{secondary} CRs produced by nuclear spallation of \textit{primary} CRs.}
\end{marginnote}

The observed linear evolution of Be with [Fe/H] is equivalent to a constant abundance ratio for the entire period of Galactic evolution: ${\rm Be/Fe} = 7.2 \times 10^{-7}$ (\textbf{Figure~\ref{fig:be9data}a}). For [Fe/H] up to about $-1$, the bulk of Fe is thought to be produced in core-collapse SNe, as Type Ia SNe do not yet contribute significantly, with a mean Fe yield per core-collapse SN of $\sim 0.07~M_\odot$, independent of the metallicity of the massive progenitor star \cite{woo95}. It implies that the Be production rate was essentially constant in the early ages of the Galaxy, with a mean Be production yield of $1.1 \times 10^{48}$ atoms per SN. This result is consistent with the predicted Be production at the current epoch, assuming that about 10\% of the total energy in SN ejecta is converted to GCR energy \cite[][see also Section 3.2.1 below]{ram97,ram00}. But the Be production yield was expected to be much lower in the early Galaxy, at the time where both the ISM and the GCRs were presumably strongly depleted in CNO nuclei. As first suggested by Duncan et al. \cite{dun92}, the observed Be evolution can be explained if GCRs, or at least the Be-producing CRs, have always had the same composition in CNO isotopes. The various astrophysical models proposed in the literature for the origin of such a CR population are discussed in Section~4. 

\textbf{Figure~\ref{fig:be9data}b} shows the same Be data as those of panel (a), but as a function of [O/H]. Oxygen is expected to provide a more direct indicator of the Be nucleosynthesis than does Fe, because Be is mainly produced by spallation of O, along with C and N. We see in \textbf{Figure~\ref{fig:be9data}b} that the Be abundances do not correlate with O as well as with Fe. The slope of $\log({\rm Be/H})$ versus [O/H] increases with increasing O abundance, which comes from the observed increase of [O/Fe] with decreasing [Fe/H] \cite[see][and references therein]{pra12b}. Fitting the sum of two linear functions of slope one and two to these logarithmic data, we find that the primary production process dominates the Be synthesis up to ${\rm [O/H]} \sim -0.5$, which corresponds approximately to the Galactic halo phase during the first billion years of Galaxy evolution \cite[see, e.g.,][]{mic13}. With a mean O yield per core collapse SN of $\sim 1.2~M_\odot$ \cite{woo95}, and the best fit ratio ${\rm Be/O} = 1.0 \times 10^{-8}$ (\textbf{Figure~\ref{fig:be9data}b}), we find that the Be production yield at that time was about $9.0 \times 10^{47}$ atoms per SN, consistent with the previous estimate based on the Fe data. At the time of solar system formation 4.6~Gyr ago, and even more so at the current epoch, Be is mainly produced as a secondary element.  

The observation that [O/Fe] increases with decreasing [Fe/H] led Fields \& Olive \cite{fie99a,fie99b} to conclude that the standard GCR nucleosynthesis model, where CRs are accelerated out of the average ISM, could explain the LiBeB origin, because the enhanced O abundance in the early Galaxy should lead to a corresponding increase of the Be production. But the data of Be vs. O still suggest the existence of a primary LiBeB production process at work in the early Galaxy (\textbf{Figure~\ref{fig:be9data}b}), which is not explained by the standard model where the GCR nuclei are drawn from the average ISM \cite[][see also Section 3.2.1 below]{ram97,ram00}.

The B abundances are also found to increase linearly with [Fe/H] \cite[see Figure 14 in][]{pra12b}, which is consistent with a constant abundance ratio: ${\rm B/Fe} = ({\rm B/Fe})_\odot = 2.2 \times 10^{-5}$. This ratio is about 30 times higher than the observed Be/Fe ratio, whereas calculations of spallogenic nucleosynthesis in the ISM by standard GCRs (Section~3.2.1) predicts the B/Be yield ratio to be $Q_{\rm B} / Q_{\rm Be} \approx 15$ \cite{ram97}. Moreover, these calculations yield an isotopic ratio $^{11}{\rm B}/^{10}{\rm B} \approx 2.5$, which is significantly lower than the ratio measured both in meteorites, $4.05 \pm 0.16$ \cite{cha95}, and in the current epoch ISM, $3.4 \pm 0.7$ \cite{lam98}. All of these results can be accounted for by a significant production of $^{11}$B by $\nu$ spallation of $^{12}$C in core-collapse SNe \citep{woo90,nak10}, which is a primary production process. However, the $^{11}$B yields of core-collapse SNe strongly depend on the neutrino temperature and total neutrino energy, and they are quite uncertain \cite[see][]{yos08}. 

\subsection{Spallogenic nucleosynthesis of light elements}
\subsubsection{Spallogenic nucleosynthesis in the ISM}

\begin{figure}[t]
\includegraphics[width=1.0\textwidth]{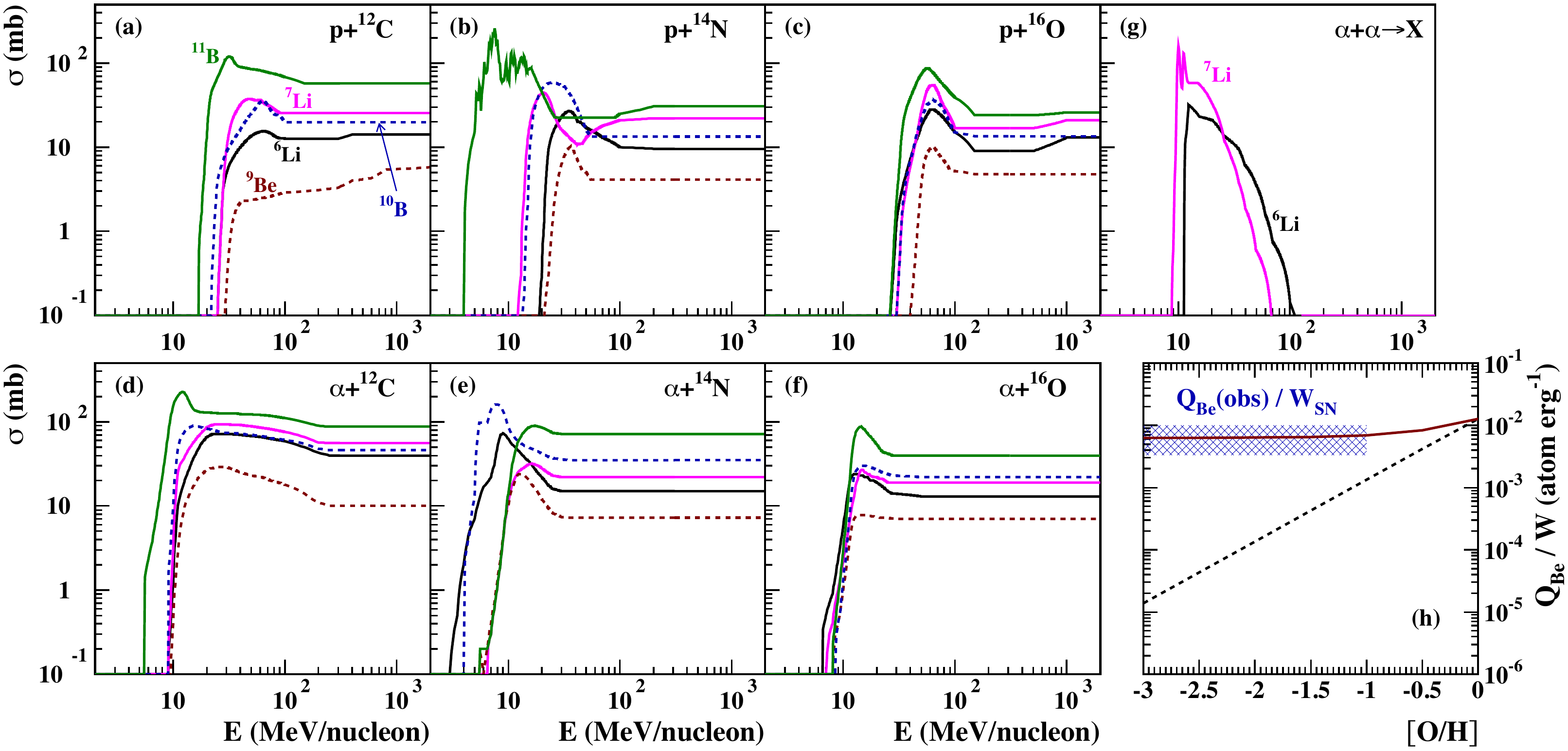}
\caption{Panels (a)--(f): cross sections for the production of $^6$Li (black solid lines), $^7$Li (magenta solid lines), $^9$Be (red dashed lines), $^{10}$B (blue dashed lines), and $^{11}$B (green solid lines) by proton and $\alpha$-particle interactions with $^{12}$C, $^{14}$N, and $^{16}$O. Data are mainly from Ref.~\cite{rea84}, augmented at high energies with data compiled in Ref.~\cite{ram97}. Panel (g): cross sections for the production of $^6$Li and $^7$Li by reactions of accelerated $\alpha$-particles with ambient He, from Refs.~\cite{rea84} and \cite{mer01}. Panel (h): Be production yield normalized to a total kinetic energy of 1 erg injected by the CR particles into the ISM, as a function of [O/H]. The two curves are obtained from the leaky box model that correctly describes the recent \textit{Voyager 1} and AMS-02 data (see Section~2.2.2 and \textbf{Figure~\ref{fig:gcrspectra}}). The red solid curve shows the case of a constant (i.e. metallicity independent) GCR composition, whereas the black dashed curve is for a time-dependent composition with CNO abundances following the evolution of [O/H] in the ISM. The blue hatched area corresponds to the normalized Be yield implied by the Be vs. O observations, assuming a CR acceleration efficiency of $1.5 \times 10^{50}$~erg per SN (see text).}
\label{fig:siglibeb}
\end{figure}

The production rate of a light isotope $l$ ($^6$Li, $^7$Li, $^9$Be, $^{10}$B and $^{11}$B) by spallogenic nucleosynthesis in the ISM can be calculated by 
\begin{equation}
Q_l = \sum_{ij} n_j \int_0^{\infty} dE v \sigma_{ij}^l(E) N_i(E) P_l(E_l,0)~,
\label{eq:y}
\end{equation}
where $i$ and $j$ range over the accelerated and ambient particle species that contribute to the synthesis of the isotope $l$, $\sigma_{ij}^l$ is the cross section for the nuclear reaction  $i+j \rightarrow l$ (see \textbf{Figure~\ref{fig:siglibeb}a-g}), $N_i$ is the differential number density of CRs of type $i$ in the ISM (see Equation~\ref{eq:n}) and $P_l(E_l,0)$ the survival probability of the freshly synthesized isotope $l$ as it slows down to rest in the ISM from its initial kinetic energy $E_l$ (see Equation~\ref{eq:p}). We have $E_l \approx E$ for the spallation reactions induced by fast CNO, because the LiBeB isotope $l$ is then produced at nearly the same energy per nucleon as that of the projectile at interaction. Reactions of fast protons and $\alpha$-particles with ambient CNO produce light isotopes of much lower initial energies, and for these direct reactions $P_l(E_l,0) \approx 1$. Inserting Equation~\ref{eq:n} into Equation~\ref{eq:y}, and noting that the energy loss rate is proportional to the ambient H density, we see that the production yields $Q_l$ do not depend on the ambient medium density, but only on the abundances $n_j/n_{\rm H}$. 

Here, we estimate the production rate of the LiBeB isotopes in the Galaxy using the GCR source parameters found to provide a good description of the recent \textit{Voyager 1} and AMS-02/HEAO-3-C2 data in the leaky box model (Section~2.3.2). %In particular, we adopt broken power laws in momentum for the source spectra $\dot{Q}_i(E)$ (equation~\ref{eq:q}), with the best-fit parameters $E_{\rm break}=200$~MeV/nucleon, $s_{\rm h.e.}=2.3$, $s_{\rm l.e.}=1.75$ for $i \equiv p$ and $\alpha$, and $s_{\rm l.e.}=1.0$ for $i \equiv {\rm CNO}$. 
\textbf{Figure~\ref{fig:siglibeb}h} shows the evolution of the normalized Be yield $Q_{\rm Be}/W$ as a function of [O/H], where 
\begin{equation}
W = \sum_{i} A_i \int_0^{\infty} dE E \dot{Q}_i(E)~
\label{eq:w}
\end{equation}
is the total power contained in the accelerated particles (the mass number $A_i$ in this Equation comes from the fact that $E$ is expressed in units of energy per nucleon). Two cases are considered for the accelerated ion composition: (\textit{i}) the GCR composition obtained from our leaky box analysis of the \textit{Voyager 1} and AMS-02 data (Section~2.3.2), assumed to be independent of [O/H] (red solid line in \textbf{Figure~\ref{fig:siglibeb}h}) and (\textit{ii}) the same GCR composition but with the abundances of the CNO nuclei scaled with ${\rm (O/H)/(O/H)}_\odot$ (black dashed line). The latter composition is representative of the assumption that the GCRs are accelerated directly out of the average ISM. For the ambient medium composition, we used the solar abundances of Ref.~\cite{lod09}, with again a scaling with ${\rm (O/H)/(O/H)}_\odot$ for the CNO elements. We also corrected the He solar abundance by a slightly decreasing function with decreasing metallicity \cite[see equation 2 in][]{par00} to take into account the He Galactic enrichment. 

The blue hatched area in \textbf{Figure~\ref{fig:siglibeb}h} shows the Be production yield per SN deduced from observations of Be abundances in metal-poor halo stars ($1.0 \times 10^{48}$ atoms; Section~3.1), divided by the total CR energy per SN, $W_{\rm SN}=1.5 \times 10^{50}$~erg \cite[i.e. 10\% of the SN total kinetic energy, see, e.g.,][]{woo95}. We have considered an error of $\pm 50$\% on $Q_{\rm Be}{\rm (obs)}/W_{\rm SN}$ to take into account the uncertainties in both the Be/O ratio (\textbf{Figure~\ref{fig:be9data}}) and the mean O yield from core-collapse SNe. As shown in \textbf{Figure~\ref{fig:siglibeb}h}, the assumption that the GCR composition in the early Galaxy was the same as today is in good agreement with the data\footnote{We note that neglecting the GCR catastrophic losses (i.e. assuming $\tau_i^{\rm tot} = \tau_i^{\rm esc}$ in Equation~\ref{eq:p}) and taking $\Lambda_{\rm esc}=10$~g~cm$^{-2}$ independent of rigidity, as done in previous LiBeB calculations \cite[e.g.][]{pra12b} increases the $Q_{\rm Be}/W$ values by a factor of 1.8--2.5 depending on [O/H].}. The inflection of the red curve for ${\rm [O/H] > -1}$ is due to the increasing contribution of the direct spallation reactions as the ISM gets enriched in CNO elements. 

On the other hand, as originally put forward by Ramaty et al. \cite{ram97,ram00}, the model where the GCRs originate from the average ISM at all times cannot explain the Be data for ${\rm [O/H] < -1}$, because of the energetic inefficiency of the Be production when both the ISM and the GCRs are depleted in CNO nuclei.
Thus, the standard model for the origin of GCRs underestimates the Be production rate at ${\rm [O/H] = -2}$ by a factor of about 50.

The GCR-produced LiBeB isotopic ratios $^7{\rm Li}/^6{\rm Li}$ and $^{11}{\rm B}/^{10}{\rm B}$ depend mainly on the nuclear cross sections and weakly on the CR spectrum, as long as the latter is not cut off at low energies ($< 100$~MeV/nucleon), where the differences in reaction thresholds can have significant effects (see \textbf{Figure~\ref{fig:siglibeb}a-g}). Thus, the isotopic ratios found in the present model, e.g. at solar metallicity $^7{\rm Li}/^6{\rm Li}=1.6$ and $^{11}{\rm B}/^{10}{\rm B}=2.5$, are in fair agreement with previous (i.e. pre-\textit{Voyager}) LiBeB calculations \cite{men71,pra93a,van00,ram00,pra12b,ram97,abi88}. 

\subsubsection{Spallogenic nucleosynthesis in supernova remnants}

Synthesis of {\it primary} Be is expected to occur in every SNR producing CRs, because before being released in the ISM at the end of the Sedov-Taylor phase, the particles accelerated at the blast wave can interact within the remnant with SN ejecta enriched in freshly synthesized CNO nuclei  \cite{par99a,par99b}.  A simple estimate of this Be production was obtained in Ref.~\cite{tat11} by assuming that a constant fraction $\theta_{\rm CR}$ of the available mechanical power processed by the blast wave is continuously transformed into kinetic energy of CRs, that can produce Be while being advected with the plasma downstream the shock front. Taking the medium inside the SNR to be well-mixed, and neglecting the adiabatic and Coulomb energy losses of the fast particles \cite[for a full calculation taking into account these losses, see][]{tat14}, the Be production rate per O-atom in the downstream plasma within the remnant can be written as 
\begin{equation}
(d{\rm Be}/dt)/{\rm O}=q_{\rm Be} \epsilon_{\rm CR}~,
\label{eq:besnr1}
\end{equation}
where $q_{\rm Be} \lsim 10^{-13}$~Be~s$^{-1}$~(O-atom)$^{-1}$~(erg~cm$^{-3}$)$^{-1}$ is the Be production rate per ambient O-atom and unit energy density in the CR particles \cite{tat11} and $\epsilon_{\rm CR}$ is the mean CR energy density in the downstream plasma. Identifying the latter to its post-shock value, we have 
\begin{equation}
\epsilon_{\rm CR} \approx k\theta_{\rm CR}\rho_{\rm CSM}u_s^2~,
\label{eq:besnr2}
\end{equation}
where $u_s$ is the forward shock velocity, $\rho_{\rm CSM}$ the mass density of the circumstellar medium (CSM) surrounding the SNR and $k$ a factor of order unity that depends on the equation of state of the shocked gas. Integrating the Be production rate (Equation~\ref{eq:besnr1}) over the SNR lifetime until the end of the Sedov-Taylor phase, we get \cite{tat11}: 
\begin{equation}
{{\rm Be} \over {\rm O}} \lsim 7 \times 10^{-10} \bigg({\theta_{\rm CR} \over 0.3}\bigg) \bigg({n_{\rm CSM} \over 1~{\rm cm}^{-3}}\bigg)^{0.64}~. 
\end{equation}
This result can be directly compared to the observed abundance ratio in low-metallicity stars, $({\rm Be}/{\rm O})_{\rm obs} \approx 1.0 \times 10^{-8}$ (\textbf{Figure~\ref{fig:be9data}}), because O in the ISM is thought to be a pure product of core-collapse SNe. Thus, to account for the Be data, a high CSM density would be needed for all SNRs, $n_{\rm CSM} \sim 100$~cm$^{-3}$, which is not realistic. Given that most SNe explode in superbubbles of low-density gas (see Sects.~2.2 and 4.3), we should have instead $n_{\rm CSM} \lsim 0.1$~cm$^{-3}$, such that spallogenic nucleosynthesis in SNRs is expected to contribute no more than a few percent to the production of primary Be in the early Galaxy. This robust conclusion is in good agreement with the results of the more detailed calculations of Refs.~\cite{par99a,par99b,tat14}. Moreover, a SNR currently producing Be at the level of $({\rm Be}/{\rm O})_{\rm obs}$ would shine in $\gamma$-rays $> 100$~MeV with a luminosity $L_\gamma \approx 10^{39}~\gamma$/s, which is significantly above the observed $\gamma$-ray luminosities of SNRs in the Galaxy \cite{fel94}. 

\section{ORIGIN OF COSMIC RAYS PRODUCING PRIMARY BERYLLIUM}

The production of \textit{primary} Be in the early Galaxy cannot be explained by the standard model for the origin of GCR nuclei, where these particles are accelerated out of the average ISM (Section~3.2.1), nor by the spallogenic synthesis of primary Be in SNRs (Section~3.2.2). The evolution of B also shows a primary behavior at low metallicity, but this could be explained by $\nu$-induced nucleosynthesis in core-collapse SNe. However, the  observed Be evolution provides new information on the nature of the reservoir(s) of material from which the particles are accelerated to become CRs. We now discuss various CR models proposed in the literature to account for the LiBeB abundances measured in metal-poor halo stars. 

\subsection{Acceleration of freshly synthesized supernova ejecta?}

As Prantzos et al. \cite{pra93b} first discussed (and criticized), the most straightforward solution to explain the Be data would be that each SN accelerates its own freshly-synthesized ejecta. Thus, elaborating on earlier ideas by, e.g., Cesarsky \& Bibring \cite{ces81}, Lingenfelter et al. \cite{lin98} suggested that high-velocity dust grains formed in SN ejecta could be the injection source for the bulk of the CR refractory elements, including C and O. In this model, the grain material is accelerated both by the SN reverse shock, when it moves back through the ejecta, and the forward shock, when high-velocity grains catch up with the slowing blast wave. Being slightly charged, these grains can be efficiently accelerated by the shocks to energies of $\sim 100$~keV/nucleon, where the friction on the background plasma sputters off individual suprathermal ions that can be further accelerated to CR energies \cite[see also][]{mey97,ell97}.

The model of Lingenfelter et al. \cite{lin98} was partly motivated by the observation of a significant broadening of the Galactic 1.809~MeV line from $^{26}$Al decay \cite{nay96}, which suggested the existence of interstellar dust grains (aluminium is a refractory element) propagating at high speed, $v_{\rm grain}\gsim 450$~km/s, some $10^6$~yr after their formation ($^{26}$Al half-life $T_{1/2}=7.2\times 10^5$~yr). But this measurement was not confirmed by subsequent gamma-ray observations with ESA's INTErnational Gamma-Ray Astrophysics Laboratory \cite[INTEGRAL;][]{die06}, and there is no more evidence that high-velocity SN grains can overtake the forward shock and escape into the ISM. Dust grains forming in the ejecta are likely processed by the reverse shock, but the energy contained in this shock is much less than the energy available in the forward shock \cite[e.g.][]{zir10}. Several other arguments have been raised that argue against this scenario \cite{mey97,mey99,ell99}, the most important being:
\begin{itemize}
\itemsep0em 
\item The low mass of the ejecta (typically $10~M_\odot$ for a core-collapse SN) compared to the mass of ISM and CSM material swept-up by the forward shock, $M_{\rm swept-up} \sim 10^3~M_\odot$ \cite[see][]{ell99}. Thus, for the bulk of the CR refractory elements to come from fresh SN ejecta and not from the ISM/CSM processed by the forward shock, the acceleration efficiency at the reverse shock must be enhanced by a factor $\sim 100$ compared to that at the forward shock, which is contrary to the observations \cite[see, e.g.,][]{zir10}. 
\item The presence in GCRs of main s-process elements such as barium, which are mainly synthesized by the slow neutron capture in low-mass stars during the asymptotic giant branch (AGB) phase. These elements are not expected to be present in significant amounts in SN ejecta, but they are found in about solar proportions in the GCR composition \cite{mey97}. 
\item The time between nucleosynthesis and acceleration, as measured from the $^{59}$Ni abundance in the GCR flux. This radioisotope decays by electron capture with a mean life in the laboratory of $1.1\times 10^5$~yr, but its decay is suppressed once the nuclei are accelerated and fully stripped. Thus, $^{59}$Ni nuclei should be observed in CRs if they are accelerated less than $\sim 10^5$~yr after the nucleosynthesis in the SN explosion \cite{cas75}, which is not the case \cite{isr05}. However, recent nucleosynthesis calculations find less $^{59}$Ni production in massive star explosions than before, which makes this argument less compelling \cite{ner16}. 
\end{itemize}
An additional argument against the SN ejecta acceleration model comes from the enrichment of lanthanides and actinides found in the GCRs \cite{don12,ale16}. This was previously considered as providing support to the ejecta model, but these elements are now thought to be produced by the r-process in binary neutron star mergers \cite{pia17}, not in core-collapse SNe. 

\subsection{Acceleration of CNO-rich stellar winds lost by supernova progenitor stars prior to explosion?}
\label{sec:gcrmodelwind}

All isotopic abundance ratios of the GCR source composition are found to be consistent with the solar composition, except the $^{22}$Ne/$^{20}$Ne ratio, which is measured to be $5.3 \pm 0.3$ times the solar value, and maybe the $^{58}$Fe/$^{56}$Fe ratio, which is estimated to be $1.69 \pm 0.27$ times solar \cite{bin08}. The large overabundance of $^{22}$Ne in the GCRs suggests a significant contribution of Wolf-Rayet (WR) star winds enriched in He-burning products (mainly $^{12}$C, $^{16}$O and $^{22}$Ne) \cite{cas82}. Thus, in the standard GCR model of Meyer, Ellison \& Drury \cite{mey97,ell97}, the $^{22}$Ne excess is accounted for by the acceleration of WR wind material when the blast waves from the most massive SNe expand in the winds lost by the progenitor massive stars prior to explosion. This scenario was studied quantitatively for the first time by Prantzos \cite{pra12a}, using models of the nucleosynthesis and evolution of both rotating and non-rotating massive stars with mass loss. He found that the observed GCR $^{22}$Ne/$^{20}$Ne ratio can be explained if the CRs are accelerated only during the early Sedov-Taylor phase, when the forward shocks run through the presupernova winds, and hardly after that, when the SN shocks propagate in the ISM. Using the shock velocity as the parameter controlling the acceleration efficiency, Prantzos \cite{pra12a} derived a minimum velocity below which the particle acceleration must be significantly suppressed: $u_s=1900$~km/s for the rotating star models and $u_s=2400$~km/s for the non-rotating ones. 

The same year, Prantzos \cite{pra12b} reassessed the problem of the LiBeB origin in the light of this new GCR model. He used massive star evolution models from the Geneva Observatory group \cite{hir05}, which show that for rotating stars, the CNO abundances in the stellar winds are almost independent of the star initial metallicity \cite{hir06}. So in this model, the GCR flux of CNO nuclei was almost the same in the early Galaxy as today, and the spallation of these nuclei in the ISM produced primary Be. Thus, introducing this new scheme for the origin of GCRs into a detailed Galactic evolution model, Prantzos \cite{pra12b} was able to provide a self-consistent description of all observational data relevant to the LiBeB production and evolution. However, several arguments make this GCR model questionable: 
\begin{itemize}
\itemsep0em 
%\item The observational signatures of on-going CR acceleration in gamma-ray bright, middle-age SNRs like W28 (age: 35--45 kyr) and W30 (age: 10--50 kyr) \cite[see][]{cap11}. In these objects, the forward shock propagates in the ISM beyond the progenitor wind zone, with a velocity $V_s \lsim 1000$~km~s$^{-1}$. The  GeV and TeV gamma-ray emissions observed from these SNRs can be explained by the collisions of accelerated nuclei with ISM particles \cite[see Section 2 and Refs.][]{cap11,gab17}.
\item The low mass contained in massive star winds compared to the total mass of material swept-up by the forward shock during the Sedov-Taylor phase. Using the Geneva Observatory database of stellar evolution models with rotation\footnote{See \url{http://obswww.unige.ch/Recherche/evol/Geneva-grids-of-stellar-evolution}}, and averaging the stellar wind masses from this database over a Salpeter initial mass function \cite[IMF; see, e.g.,][]{pag}, we find that the mean mass of wind material lost by massive stars (of initial mass between $10$ and $120~M_\odot$) prior to explosion is $M_{\rm wind} \approx 12~M_\odot$, which is about two orders of magnitude lower than $M_{\rm swept-up}$ \cite{ell99}. Thus, the model requires a similar enhancement in the accelerated population of wind material over swept-up ISM material, which is not predicted by diffusive shock acceleration theory. In Reference~\cite{pra12a}, the selective acceleration of wind material is achieved due to the assumed minimum speed of $u_s \approx 2000$~km/s for the acceleration process to be effective.  
\item The small amount of energy lost by the blast wave during the early stage of propagation through the progenitor wind. Adopting the model of Ref.~\cite{pra12a} for the SN shock evolution into the CSM, we find that the forward shock looses only about 10\% of its total energy during this early phase. This raises a potential issue with the requirement that the GCRs must acquire $\sim 10$\% of the total kinetic energy of the SN ejecta (Section~2). 
\item The contribution of thermonuclear SNe (Type Ia) to the GCR production. The blast waves from these SNe are rapidly expanding into the average ISM (i.e. there is no wind zone in the CSM) and several remnants from Type Ia SNe (e.g. SN~1006 and Tycho's SNR) show signatures of efficient CR acceleration \cite{hel12}. Assuming that the CR acceleration process stops at the end of the Sedov-Taylor stage for remnants of SNe~Ia, but that it stops at the boundary of the stellar wind bubble for remnants of core-collapse SNe, the $^{22}$Ne/$^{20}$Ne isotopic ratio in the GCRs can be estimated as:
\begin{equation}
\bigg(\frac{^{22}{\rm Ne}}{^{20}{\rm Ne}}\bigg)_{\rm GCR} \approx \frac{(^{22}{\rm Ne}/^{20}{\rm Ne})_{\rm wind}M_{\rm wind}+(^{22}{\rm Ne}/^{20}{\rm Ne})_{\rm ISM}M_{\rm swept-up}f_{\rm SNIa}}{M_{\rm wind}+M_{\rm swept-up}f_{\rm SNIa}}~,
\label{eq:22ne}
\end{equation}
where $(^{22}{\rm Ne}/^{20}{\rm Ne})_{\rm wind}\approx 0.63$ is the IMF-averaged Ne isotopic ratio in massive star winds from the stellar yields of the Geneva group \cite{hir05}, $M_{\rm wind} \approx 12~M_\odot$ (see above), $(^{22}{\rm Ne}/^{20}{\rm Ne})_{\rm ISM}=7.35 \times 10^{-2}$ is the solar abundance ratio \cite{lod09}, $M_{\rm swept-up} \sim 10^3~M_\odot$ \cite{ell99}, and $f_{\rm SNIa}\approx 0.25$ is the ratio of thermonuclear to core-collapse SN rates in the current epoch Galaxy. Equation~\ref{eq:22ne} gives $(^{22}{\rm Ne}/^{20}{\rm Ne})_{\rm GCR}\approx 0.1$, which is significantly less than the observed GCR $^{22}{\rm Ne}/^{20}{\rm Ne}$ ratio \cite{bin08}: $0.387 \pm 0.007$ (statistical) $\pm 0.022$ (systematic). 
\item The abundances of the refractory elements in the GCR source composition, which are in solar proportions (to within 20\%). Mg, Si and Ca are mainly produced in core-collapse SNe, whereas $\sim 70$\% of Fe, Co and Ni is currently synthesized in SNe~Ia \cite[][]{tim95}. Once created, these refractory elements are thought to be rapidly locked in dust grains formed in SN ejecta. The fact that they are found in solar proportions in the GCR source composition provides strong evidence that the accelerated particles come from various dust grains of the ISM mix \cite{mey97}. The chemical nature of the dust grains formed in hot stellar winds, such as those of carbon-rich WR stars, is markedly different from that of the dust forming in other astrophysical environments \cite{che10}. 
\item The observational constraints on the stellar progenitors of core-collapse SNe that suggest that most, if not all, WR stars do not end their life in SN, but collapse to form a black hole \cite{sma15}. These objects would thus not produce a shock wave that can accelerate the $^{22}$Ne nuclei contained in the wind from the progenitor star. 
\end{itemize}
For all these reasons, the massive star wind acceleration model seems unlikely. 

\subsection{Acceleration of cosmic rays in superbubbles?}
\label{sec:SBLiBeB}

%Massive stars are mainly found in OB associations containing up to several tens of OB stars, and have relatively short lifetimes (about 3 to 20 Myr), such that the majority of core-collapse SNe go off in hot, low-density, wind-generated superbubbles. 

In the superbubble model for the origin of CRs, the energetic particles are accelerated out of a mix of average ISM material and fresh ejecta of massive star winds and core-collapse SNe within hot, low-density superbubbles generated by the stellar activity of OB associations (Section~2.2). In this scenario, CNO nuclei synthesized by massive stars from an OB association can be accelerated by SN shocks from the explosion of stars from the same cluster, such that the subsequent spallation of these CNO nuclei in the ISM can produce primary Be  \cite{par99c}. Shock acceleration is expected to be most effective in hot and low-density superbubbles \cite{axf81}, which are in fact the ``hot ionized phase'' of the ISM. As discussed in Section~2.2, repeated shock acceleration of low-energy particles in the superbubble environment can produce a hardening of the CR source spectrum in the non-relativistic energy range, and the predicted CR spectrum may be consistent with the \textit{Voyager} data (see Section~2.3.2).  

A potential issue with the superbubble model arises from the $^{59}$Co and $^{59}$Ni observations in the GCR flux, which suggest a delay of $\sim 10^5$~yr between SN nucleosynthesis and acceleration \cite{isr05}. According to Higdon \& Lingenfelter \cite{hig98}, the mean time between successive SNe in superbubbles is $\sim 3 \times 10^5$~yr, which is long enough to allow for the decay of the bulk of $^{59}$Ni between two SN explosions. But Prantzos \cite{pra05} noted that $^{59}$Ni may be continuously accelerated in superbubbles by the plasma turbulence excited by strong stellar winds \cite[see also][]{par04}. However, revised calculations for the stellar yields of $^{59}$Co and $^{59}$Ni essentially removes the requirement of $^{59}$Ni decay prior to acceleration \cite{ner16}.

The composition of the matter within superbubbles is not well determined either observationally nor theoretically. In the classical superbubble model \cite{mac88}, the bulk of the hot superbubble gas is provided by conductive evaporation from the cold outer shell of the swept-up ISM. The mass in SN ejecta and stellar wind material then accounts to only a few percents of the total superbubble mass \cite[see][]{hig98}. But the cores of superbubbles, where most SNe occur, could be more enriched in fresh ejecta than the outer zones, depending on the level of gas mixing in the superbubble interior. It is also possible that the heating and evaporation of the outer shell are suppressed by swept-up interstellar magnetic fields, which would significantly increase the relative abundance of fresh ejecta in the hot gas \cite{hig98}. According to Lingenfelter \& Higdon \cite{lin07a}, the average mass fraction of SN and WR ejecta in the SN-active cores of superbubbles is about 17\%$\pm$5\%, which corresponds to a superbubble core metallicity of $\sim 2.1$ times that of the average ISM at present epoch, given that the IMF-averaged metallicity of SN and massive star wind ejecta is about 7.5 times the present ISM value (and about 10 times the protosolar metallicity). Such a high metallicity of superbubble gas may not be supported by soft X-ray observations of the hot ISM, nor by that of superbubbles in the Large Magellanic Cloud \cite[see, e.g.,][]{deh14}. However, X-ray measurements may underestimate the actual metallicity of hot gas if most of the metals are locked in dust grains \cite{bal06,lin07b}.

\begin{figure}[t]
\includegraphics[width=0.65\textwidth]{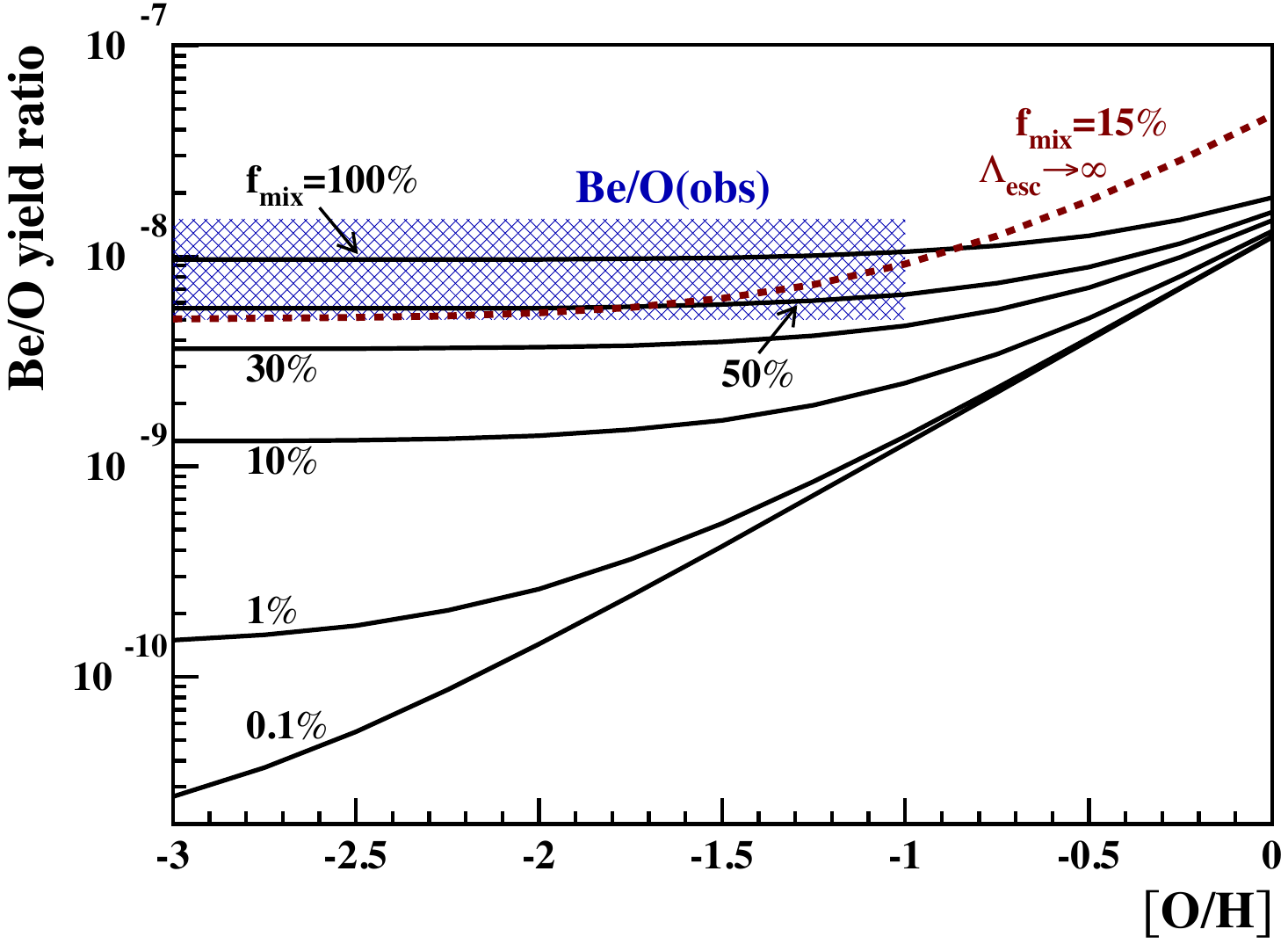}
\caption{Calculated Be/O yield ratio in the superbubble model as a function of [O/H] in the average ISM, for various values of the mixing parameter $f_{\rm mix}$ (Equation~\ref{eq:fmix}). The solid curves are for the leaky box model described in Section~2.3.1., whereas the red dashed curve is for a closed Galaxy CR propagation model (i.e. $\Lambda_{\rm esc} \rightarrow \infty$), with $f_{\rm mix}=15\%$. The blue hatched area corresponds to the Be/O ratio found in metal-poor halo stars (see \textbf{Figure~\ref{fig:be9data}b}).}
\label{fig:superbubble}
\end{figure}

In \textbf{Figure~\ref{fig:superbubble}}, we study the level of mixing of fresh SN ejecta and old ISM required to account for the Be/O ratio measured in metal-poor halo stars. In practice, we took the source abundance of CRs of type $i$ ($p$, $\alpha$, $^{12}$C, $^{14}$N, and $^{16}$O) to be 
\begin{equation}
K_i^{\rm CR} = f_{\rm mix} K_i^{\rm ej} + (1-f_{\rm mix})~K_i^{\rm ISM},
\label{eq:fmix}
\end{equation}
where $K_i^{\rm ej}$ is the abundance of isotope $i$ in the IMF-averaged SN ejecta, for which we used the yields of Ref.~\cite{woo95} for SNe of initial metallicity  $Z=0.01Z_\odot$ (normalized to $K_p^{\rm ej}=1$, we find $K_\alpha^{\rm ej}=0.195$, $K_{\rm ^{12}C}^{\rm ej}=1.8 \times 10^{-3}$, $K_{\rm ^{14}N}^{\rm ej}=5.9 \times 10^{-6}$ and $K_{\rm ^{16}O}^{\rm ej}=8.8 \times 10^{-3}$)\footnote{While the $p$, $\alpha$, C and O abundances are largely independent of metallicity, it is not the case for N, which is a secondary element. However, the contribution of N spallation to the Be production is negligible.}, and $K_i^{\rm ISM}$ is the abundance of specie $i$ in the average ISM, where, as before (Section~3.2.1), we scaled the CNO abundances with ${\rm (O/H)/(O/H)}_\odot$ and multiplied the He solar abundance with a slowly varying function of [O/H] to account for the Galactic enrichment. We also assume that the broken power-law CR source spectra obtained from the \textit{Voyager} data (Section~2.3.2) are representative of the average energy distributions of CRs in the Galaxy since its formation. 

The Be/O production (or yield) ratio is obtained from  
\begin{equation}
{{\rm Be} \over {\rm O}} = {(Q_{\rm Be}/W) W_{\rm SN} m_{\rm O} \over M_{\rm O}^{\rm ej}}~, 
\label{eq:be-to-o}
\end{equation}
where $(Q_{\rm Be}/W)$ is the normalized Be production rate per unit CR power calculated with the same leaky box model as before (see Sects.~2.3.1, 3.2.1 and \textbf{Figure~\ref{fig:siglibeb}h}), $W_{\rm SN}=1.5 \times 10^{50}$~erg is the assumed total CR energy per SN, $m_{\rm O}$ is the mass of the O atom, and $M_{\rm O}^{\rm ej}=1.2~M_\odot$ the mean mass of O produced in a core-collapse SN \cite{woo95}. 

\textbf{Figure~\ref{fig:superbubble}} shows that the calculated Be/O yield ratio evolves from a constant value at low metallicity, characteristic of a primary Be production process, to a ratio proportional to (O/H) above a transition value [O/H]$_t$, which depends on the mixing parameter $f_{\rm mix}$ \cite[see also][]{par00}. Such a behavior is consistent with the Be vs. O data (\textbf{Figure~\ref{fig:be9data}b}). As long as the Be/O yield ratio is constant, it can be directly compared to the abundance ratio measured in metal-poor halo stars (blue hatched area in \textbf{Figure~\ref{fig:superbubble}}), but for ${\rm [O/H] > [O/H]_t}$, the comparison with observations requires the use of a Galactic chemical evolution model \cite[e.g.][]{ali02}. In the leaky box model adopted to fit the \textit{Voyager 1} and AMS-02 data (Section~2.3), a mixing fraction $f_{\rm mix} \gsim 45\%$ is required to account for the observed Be/O ratio at low metallicity (\textbf{Figure~\ref{fig:superbubble}}). This would imply a superbubble core metallicity at present time $\gsim 5.2$ times solar, which is not supported neither by observations nor by theory (see above). 

However, it is likely that the GCRs were more efficiently confined in the Galaxy during its early halo phase than today, because the ISM turbulence was probably higher at that time and the halo boundary 
%(i.e. $H$ in Equation~\ref{eq:lambda}) 
was farther \cite{pra93a}. We see from \textbf{Figure~\ref{fig:superbubble}} that in a closed-Galaxy CR propagation model with $\Lambda_{\rm esc} \rightarrow \infty$, the observed Be/O ratio can be reproduced with $f_{\rm mix} \gsim 15\%$. Taking into account that in the early Galaxy $\sim 80-90\%$ of all Galactic SNe (mostly core-collapse) occurred in OB associations \cite{hig05}, the mean SN-WR ejecta mass fraction in the SN-active core of superbubbles should then be $\gsim 17\%$, which is consistent with the estimate of Lingenfelter \& Higdon \cite{lin07a}. It is also consistent with measurements of GCR metal abundances, which suggest that the CR source material consists of a mixture of $\sim 20\%$ ejecta from massive stars and $\sim 80\%$ ISM material with solar composition \cite[][and references therein]{rau09,mur16}.

Higdon \& Lingenfelter \cite{hig03} found that a similar mass fraction of $18\%\pm5\%$ of SN-WR ejecta in superbubble cores can explain the anomalously high $^{22}$Ne/$^{20}$Ne ratio of the GCR source composition (see Section~4.2). But using more recent stellar yield calculations \cite{hir05}, Prantzos \cite{pra12b} found that the Ne isotopic ratio in superbubbles should be close to solar, which is to be expected as massive stars are the main source of both $^{20}$Ne and $^{22}$Ne. About 40 years after the discovery of the GCR $^{22}$Ne excess \cite{gar79}, the origin of this important isotopic anomaly is still not explained. 

\subsection{A distinct component of low-energy cosmic rays?}

Another CR model that arose in the 1990s invokes the existence of an additional component of low-energy CRs (LECR), with kinetic energies $\lsim$ a few hundreds of MeV/nucleon, which would be responsible for the production of the bulk of primary Be \cite{cas95,van96,ram97}. This model is distinct from the ``carrot'' model proposed in the 1970s to enhance the production of $^{11}$B \cite[][see Section~3.1]{men71}, because the low-energy CRs in the latter model were supposed to have the same composition as the standard GCRs and thus would produce only secondary Be. In the model developed in the 1990s after the LiBeB abundance measurements in metal-poor halo stars, the low-energy CRs are preferentially accelerated in the winds of massive stars or in massive SN explosions within superbubbles \cite{van98,lem98}. 

However, the recent CR observations by the \textit{Voyager 1} probe beyond the heliopause \cite{cum16} seem to argue against this model, given that (i) the measured spectra do not show any additional flux of low-energy CRs, and (ii) the solar system is currently inside a small superbubble, namely the Local Bubble, which was created by the activity of about 14 to 20 massive stars originating from the Scorpius-Centaurus OB association \cite{fuc06}. 

The total CR ionization rate of atomic hydrogen resulting from the \textit{Voyager} spectra, $\zeta_{\rm H}=(1.51 - 1.64)\times10^{-17}$~s$^{-1}$, is, however, a factor $>10$ lower than the average CR ionization rate of $\zeta_{\rm H}=1.78\times10^{-16}$~s$^{-1}$ \cite{ind15} measured in clouds across the Galactic disc, using line observations of ionized molecules by the Herschel satellite \cite{neu17}. This difference may suggest that LECRs are relatively less abundant in the local ISM than elsewhere in the Galaxy. Observations of  H$_3^+$  in diffuse molecular clouds show indeed that the density of LECRs can strongly vary from one region to another in the Galactic disk, and, in particular, that LECR fluxes can be significantly higher than the average value in diffuse molecular gas residing near sites of CR acceleration such as SNRs \cite[][and references therein]{ind12,mereview}. It is not known, however, if the extra ionization is due to LECR ions or electrons \cite{pad09}. 

The best way to test the existence of additional fluxes of LECR nuclei in the ISM and probe the distribution of these particles in different Galactic environment would be to detect the characteristic gamma-ray lines between $0.1$ and $10$~MeV produced by nuclear collisions of CRs with interstellar matter \cite{ben13}. The detection of this long-predicted emission \cite{ram79} can be revealed by a future gamma-ray space mission like e-ASTROGAM \cite{eastrogam}. 

%\begin{figure}[ht]
%\includegraphics[width=0.95\textwidth]{glcem2.pdf}
%\caption{Panel (a): predicted gamma-ray emission due to nuclear interactions of CRs in the inner Galaxy (longitude $-80^\circ \leq l \leq 80^\circ$ and latitude $-8^\circ \leq b \leq 8^\circ$). The gamma-ray line emission below 10~MeV is due to LECRs, whose properties in the ISM have been adjusted such that the mean CR ionization rate deduced from H$_3^+$ observations and the  \textit{Fermi}-LAT data (magenta band) at 1 GeV are simultaneously reproduced (adapted from \cite{ben13}). The dashed green line shows the total calculated emission when adding leptonic contributions, point sources and extragalactic gamma-ray background that were taken from \cite{ack12}. Panel (b): Zoom on the nuclear line region of the gamma-ray spectrum (not multiplied by $E^2$), with indications of the emitting nuclei for some prominent lines.}
%\label{fig:lines}
%\end{figure}

\section{SUMMARY AND OUTLOOK} 

We have reviewed the SNR paradigm for the origin of GCRs in the light of the recent CR data acquired by the \textit{Voyager 1} and AMS-02 space instruments. We have pointed out that the spectrum of CRs released from SNRs is likely to show a deviation from the pure power law in momentum predicted by the diffusive shock acceleration theory for the particle distribution function at the shock, and that this deviation should take the form of a hardening of the released CR spectrum towards lower momenta. A spectral break is indeed expected at the transition between high-energy CRs continuously escaping the SNR shock from upstream during the Sedov-Taylor phase and low-energy CRs released at the end of the SNR evolution, when the shock becomes radiative and eventually dissolves in the ISM. We have also discussed the acceleration of CRs in superbubbles powered by multiple SNe correlated in both time and space. In this case also, a hardening of the CR source spectrum at low energies can be expected, due to the diffusion of these particles through multiple shocks in the superbubble interior.  

Using a standard leaky box model to describe the propagation of CRs in the ISM, we have shown that the CR energy distributions recently measured by \textit{Voyager 1} and AMS-02 are consistent with the expected spectral hardening at low energies. In particular, the measured CNO spectra can be fitted with a very hard CR source spectrum below $E_{\rm break} \approx 200$~MeV/nucleon, $f_i(p) \propto p^{-3}$, which is close to the predicted CR spectrum resulting from multiple shock acceleration in superbubbles. The derived proton and $\alpha$-particle source spectra are somewhat softer, the best-fit power-law index being $q=3.75$ below $E_{\rm break}$, and the origin of this difference definitely deserves further attention. Clearly, the recent data collected by the \textit{Voyager 1} probe beyond the heliopause still have a lot to tell us about the origin of CRs in the Galaxy. 

We have then reassessed the theory for the spallogenic origin of LiBeB isotopes in the light of the new information gained on low-energy CRs. The Be abundances measured in metal-poor halo stars also provide information on CR origin, because these data cannot be explained by the standard GCR model in which the energetic particles are accelerated out of the average ISM. Specifically, the observed Be abundances vs. [O/H] suggest that in the early halo phase of the Galaxy, Be was mainly produced by spallation of fast CNO nuclei that were much more abundant (relative to H) in the GCR composition than in the average ISM composition at that time. The superbubble model for the origin of GCRs provides a plausible explanation for these observations by arguing that the reservoir of material in superbubble cores from which the particles are accelerated, is enriched in fresh ejecta of massive star winds and core-collapse SNe. Assuming that the CR source spectra obtained from the \textit{Voyager} data are representative of the CR energy distributions in the early Galaxy, and that the GCRs were more efficiently confined in the Galaxy during its early halo phase than today, we have found that a mixture in superbubble cores of $\gsim 17$\% of ejecta from massive stars and $\lsim 83$\% of older ISM material can account for the observed Be abundances in low-metallicity stars. In this model, Be is mainly produced at the current epoch by spallation of interstellar CNO nuclei, i.e. as a secondary element, which is consistent with the observed Be evolution as a function of [O/H]. 
The recent very first detections of superbubbles in both the GeV and TeV energy domain \cite{latcocoon,hessLMC} encourages belief that in the near future the superbubble scenario for the origin of GCRs might become testable by means of gamma-ray observations. 

%A crucial test of this scenario might come from a simultaneous interpretation of the LiBeB data and the recent detections of superbubbles in gamma rays

%Alternatively, we cannot exclude that the bulk of primary Be in the early Galaxy was produced by an additional component of low-energy CRs... 

%Here is an example of a extract.
%\begin{extract}
%This is an example text of quote or extract.
%\end{extract}

%\subsection{Sidebars and Margin Notes}
% Margin Note
%\begin{marginnote}[]
%\entry{Metallicity}{[Fe/H]=$\log[{\rm (Fe/H)/(Fe/H)}_\odot]$, (Fe/H)/(Fe/H)$_\odot$ being the Fe abundance relative to its solar value}
%\entry{Term B}{definition}
%\end{marginnote}

%\begin{textbox}[ht]\section{SIDEBARS}
%Sidebar text goes here.
%\subsection{Sidebar Second-Level Heading}
%More text goes here.\subsubsection{Sidebar third-level heading}
%Text goes here.\end{textbox}

% Summary Points
%\begin{summary}[SUMMARY POINTS]
%\begin{enumerate}
%\item Summary point 1. These should be full sentences.
%\item Summary point 2. These should be full sentences.
%\end{enumerate}
%\end{summary}

% Future Issues
%\begin{issues}[FUTURE ISSUES]
%\begin{enumerate}
%\item Future issue 1. These should be full sentences.
%\item Future issue 2. These should be full sentences.
%\end{enumerate}
%\end{issues}

%Disclosure
\section*{DISCLOSURE STATEMENT}
The authors are not aware of any affiliations, memberships, funding, or financial holdings that might be perceived as affecting the objectivity of this review. 

% Acknowledgements
\section*{ACKNOWLEDGMENTS}
The authors acknowledge support from Agence Nationale de la Recherche (grant ANR-17-CE31-0014), and would like to thank E. Amato, W. R. Binns, L. Drury, J. Kiener, A.  Marcowith, G. Morlino, E. Parizot, N. Prantzos, and E. Vangioni for fruitful discussions. V.T. is grateful to ISSI (International Space Institute at Bern) for the support and hosting of the International Team \#351 on Galactic Cosmic Ray Origin and Composition.


\begin{thebibliography}{96}
\expandafter\ifx\csname
natexlab\endcsname\relax\def\natexlab#1{#1}\fi

%%INTRODUCTIONVVVVVVVVVVVVVVVVVVVV

\bibitem{blasireview}
Blasi, P. \textit{Astron. Astrophys. Rev.} 21:70 (2013)

%\bibitem{caprioliICRC} Caprioli, D. arXiv:1510.07042

\bibitem{lukerecentreview} Drury, L. O'C., \textit{Proc. 35th ICRC}, arXiv:1708.08858

\bibitem{hel12} Helder, E.~A., Vink, J., Bykov, A.~M., et al. \textit{Space Science Rev.} 173:369 (2012) 

\bibitem{fermipion} Ackermann, M., Ajello, M., Allafort, A., et al.\ 2013, Science, 339, 807 

\bibitem{latsnrcat} Acero, F., Ackermann, M., Ajello, M., et al. \textit{Astrophys. J. Suppl.} 224:8 (2016)

\bibitem{latcocoon} Ackermann, M., Ajello, M., Allafort, A., et al. \textit{Science} 334:1103 (2011)

\bibitem{latdiffuse} Ackermann, M., Ajello, M., Atwood, W.~B., et al.  \textit{Astrophys. J.} 750:3 (2012)

\bibitem{gre15} Grenier, I.~A., Black, J.~H., \& Strong, A.~W. \textit{Ann. Rev.Astron. Astrophys.} 53:199 (2015)

\bibitem{ree70} Reeves, H., Fowler, W. A., \& Hoyle, F. \textit{Nature} 226:727 (1970)

\bibitem{men71} Meneguzzi, M., Audouze, J., \& Reeves, H. \textit{Astron. Astrophys.} 15:337 (1971)

\bibitem{van90} Vangioni-Flam, E., Audouze, J., Oberto, Y., \& Casse, M. \textit{Astrophys. J.} 364:568 (1990)

\bibitem{pra93a} Prantzos, N., Casse, M., \& Vangioni-Flam, E. \textit{Astrophys. J.} 403:630 (1993)

\bibitem{van00} Vangioni-Flam, E., Cass{\'e}, M., \& Audouze, J. \textit{Phys. Rep.} 333:365 (2000)

\bibitem{ram00} Ramaty, R., Scully, S.~T., Lingenfelter, R.~E., \& Kozlovsky, B. \textit{Astrophys. J.} 534:747 (2000)

\bibitem{pra12b} Prantzos, N. \textit{Astron. Astrophys.} 542:A67 (2012) 

\bibitem{sto13} Stone, E.~C., Cummings, A.~C., McDonald, F.~B., et al. \textit{Science} 341:150 (2013)

\bibitem{cum16} Cummings, A.~C., Stone, E.~C., Heikkila, B.~C., et al. \textit{Astrophys. J.} 831:18 (2016)

\bibitem{ams1} Aguilar, M., Aisa, D., Alpat, B., et al. \textit{Physical Review Letters} 114:171103 (2015)

\bibitem{ams2} Aguilar, M., Aisa, D., Alpat, B., et al. \textit{Physical Review Letters} 119:251101 (2017)

\bibitem{ind12} Indriolo, N., \& McCall, B.~J. \textit{Astrophys. J.} 745:91 (2012)

\bibitem{neu17}  Neufeld, D.~A., \& Wolfire, M.~G. \textit{Astrophys. J.} 845:163 (2017)

%%%%%%%%%%%%VVVVVVVVVVVVVVVVVVVV

\bibitem{hillas}
Hillas, A.M. \textit{J. Phys. G: Nucl. Part. Phys.} 31:R95 (2005)

\bibitem{baadezwicky}
Baade, W., Zwicky, F. \textit{Proc. Nat. Ac. Sci.} 20:259 (1934)

\bibitem{terhaar}
ter Haar, D \textit{Rev. Mod. Phys.} 22:119 (1950)

\bibitem{strongCRpower}
Strong, A. W., Porter, T. A., Digel, S. W., et al. \textit{Astrophys. J.} 722:L58 (2010)

%\bibitem{krymskii} Krymskii, G. F. \textit{Dokl. Phys.} 22:327 (1977) 

%\bibitem{axford} Axford, W. I., Leer, E., Skadron, G. \textit{ICRC} 11:132 (1977)

%\bibitem{blandford} Blandford, R. D., Ostriker, J. P.  \textit{Astrophys. J.} 221:L29 (1978)

\bibitem{luke83} Drury, L.O'C. \textit{Rep. Prog. Phys.} 46:973 (1983)

\bibitem{strongreview}
Strong, A.W., Moskalenko, I.V., Ptuskin, V.S. \textit{Ann. Rev. Nucl. Part. Sci.} 57:285 (2007)

\bibitem{mau10} Maurin, D., Putze, A., \& Derome, L. \textit{Astron. Astrophys.} 516:A67 (2010)

\bibitem{ellisonbowshock}
Ellison, D. C., Moebius, E., Paschmann, G. \textit{Astrophys. J.} 352:376 (1990)

\bibitem{ginzburgsyrovatskii}
Ginzburg, V. L., Syrovatskii, S. I. \textit{The origin of cosmic rays} (New York: Macmillan, 1964)

\bibitem{koyamaelectrons} 
Koyama, K., Petre, R., Gotthelf, E. V., et al. \textit{Nature} 378:255 (1995)

\bibitem{felixbook}
Aharonian, F.A. \textit{Very high energy cosmic gamma radiation : a crucial window on the extreme Universe}, 2004 (World Scientific)

\bibitem{felixSNRs}
Aharonian, F.A. \textit{Astropart. Phys.} 43:71 (2013)

\bibitem{mereview}
Gabici, S., Montmerle, T. \textit{Proc. 25th ICRC} p.29 (2015), arXiv:1510.02102

\bibitem{etienne}
Parizot, E. \textit{Nucl. Phys. B: Proc. Suppl.} 256:97 (2013)

\bibitem{felixreview}
Aharonian, F.A., Buckley, J., Kifune, T., Sinnis, G. \textit{Rep. Prog. Phys.} 71:096901 (2008)

\bibitem{klara}
Schure, K.M., Bell, A.R. \textit{Mon. Not. R. Astron. Soc.} 435:1174 (2013)

\bibitem{damianoinjection}
Caprioli, D., Pop. A.-R., Spitkovsky, A. \textit{Astrophys. J.} 798:L28 (2015)

\bibitem{bell} Bell, A. R. \textit{Mon. Not. R. Astron. Soc.} 182:147 (1978)

\bibitem{malkov}
Malkov, M.A., Drury, L.O'C. \textit{Rep. Prog. Phys.} 64:429 (2001)

\bibitem{klarareview}
Schure, K.M., Bell, A.R., Drury, L.O'C., Bykov, A.M. \textit{Space Sci. Rev.} 173:491 (2012)

\bibitem{damianosteep}
Caprioli, D. \textit{J. Cosm. Astropart. Phys.} 07:38 (2012)

\bibitem{bell13}
Bell, A.R., Schure, K.M., Reville, B., Giacinti, G. \textit{Mon. Not. R. Astron. Soc.} 431:415 (2013)

\bibitem{meescape}
Gabici, S. \textit{Mem. Soc. Astron. It.} 82:760 (2011)

\bibitem{pz05}
Ptuskin, V.S., Zirakashvili, V.N. \textit{Astron. Astrophys.} 429:755 (2005)

\bibitem{lod09} Lodders, K., Palme, H., \& Gail, H.-P. \textit{The Landolt B{\"o}rnstein Database, New series, VI/4B, Springer, Berlin}, J.E. Tr\"umper (Ed.) (2009)

\bibitem{rau09} Rauch, B.~F., Link, J.~T., Lodders, K., et al. \textit{Astrophys. J.} 697:2083 (2009)

\bibitem{mur16} Murphy, R.~P., Sasaki, M., Binns, W.~R., et al. \textit{Astrophys. J.} 831:148 (2016)

\bibitem{mey97} Meyer, J.-P., Drury, L.~O., \& Ellison, D.~C. \textit{Astrophys. J.} 487:182 (1997)

\bibitem{ell97} Ellison, D.~C., Drury, L.~O., \& Meyer, J.-P. \textit{Astrophys. J.} 487:197 (1997)

\bibitem{wiedenbeck}
Wiedenbeck, M.E., et al. \textit{Space Sci. Rev.} 130:415 (2007)

\bibitem{cap17} Caprioli, D., Yi, D.~T., \& Spitkovsky, A. \textit{Phys. Rev. Letters} 119:171101 (2017)

\bibitem{morlino}
Morlino, G. \textit{Mon. Not. R. Astron. Soc.} 412:2333 (2011)

\bibitem{garmany}
Garmany, C.D. \textit{Publ. Astron. Soc. Pac.} 106:25 (1994)

\bibitem{hig98} Higdon, J.~C., Lingenfelter, R.~E., \& Ramaty, R. \textit{Astrophys. J.} 509:L33 (1998)

\bibitem{kaf81} Kafatos, M., Bruhweiler, F., \& Sofia, S. \textit{International Cosmic Ray Conference} 2:222 (1981)

\bibitem{mac88} Mac Low, M.-M., \& McCray, R. \textit{Astrophys. J.} 324:776 (1988)

\bibitem{byk90} Bykov, A.~M., \& Toptygin, I.~N. \textit{Zhurnal Eksperimentalnoi i Teoreticheskoi Fiziki} 98:1255 (1990)

\bibitem{byk92} Bykov, A.~M., \& Fleishman, G.~D. \textit{Mon. Not. R. Astron. Soc.} 255:269 (1992)

\bibitem{byk14} Bykov, A.~M. \textit{Astron. Astrophys. Rev.} 22:77 (2014)

\bibitem{par00} Parizot, E. \textit{Astron. Astrophys.} 362:786 (2000)

\bibitem{par04} Parizot, E., Marcowith, A., van der Swaluw, E., Bykov, A.~M., \& Tatischeff, V. \textit{Astron. Astrophys.} 424:747 (2004)

\bibitem{ferrand}
Ferrand, G., Marcowith, A. \textit{Astron. Astrophys.} 50:A101 (2010)

%%%%%%%%%%%%%%%%%%%%%%%%%%%%%%%%%%%%%%%%%%%%%

%\bibitem{cap11} Caprioli, D., Blasi, P., \& Amato, E. \textit{Astroparticle Phys.} 34:447 (2011)

%\bibitem{par99} Parizot, E., \& Lehoucq, R. \textit{Astron. Astrophys.} 346:211 (1999) 

\bibitem{web90} Webber, W.~R., Kish, J.~C., \& Schrier, D.~A. \textit{Phys. Rev.} C41:520 (1990) 

\bibitem{tri99} Tripathi, R.~K., Cucinotta, F.~A., \& Wilson, J.~W. \textit{Nucl. Instrum. Methods Phys. Res.} B155:349 (1999)

\bibitem{mos02} Moskalenko, I.~V., Strong, A.~W., Ormes, J.~F., \& Potgieter, M.~S. \textit{Astrophys. J.}  565:280 (2002)

%\bibitem{ber05} Berger, M.J., Coursey, J.S., Zucker, M.A., \& Chang, J. 2005, ESTAR, PSTAR, and ASTAR: Computer Programs for Calculating Stopping-Power and Range Tables for Electrons, Protons, and Helium Ions (version 1.2.3), National Institute of Standards and Technology, Gaithersburg, MD (Available: http://physics.nist.gov/Star) (2005)

%\bibitem{pie68} Pierce, T.~E., \& Blann, M. \textit{Phys. Rev.} 173:390 (1968)
\bibitem{schlibook} Schlickeiser, R., \textit{Cosmic ray astrophysics}, Astronomy and Astrophysics Library; Physics and Astronomy Online Library.~Berlin: Springer.~ISBN 3-540-66465-3 (2002)  

\bibitem{eng90} Engelmann, J.~J., Ferrando, P., Soutoul, A., Goret, P., \& Juliusson, E. \textit{Astron. Astrophys.} 233:96 (1990)

\bibitem{jon01} Jones, F.~C., Lukasiak, A., Ptuskin, V., \& Webber, W. \textit{Astrophys. J.}  547:264 (2001)

%\bibitem{fer98} Ferri{\`e}re, K. \textit{Astrophys. J.}  497:759 (1998)

\bibitem{B2FH} Burbidge, E.~M., Burbidge, G.~R., Fowler, W.~A., \& Hoyle, F. \textit{Rev. Modern Phys.} 29:547 (1957)

\bibitem{ree94} Reeves, H.\ 1994, \textit{Rev. Modern Phys.} 66:193 (1994)

\bibitem{spi82} Spite, F., \& Spite, M. \textit{Astron. Astrophys.} 115:357 (1982)

\bibitem{coc17} Coc, A., \& Vangioni, E. \textit{Int. J. Modern Phys.} E26:1741002 (2017)

\bibitem{coc14} Coc, A., Uzan, J.-P., \& Vangioni, E. \textit{J. Cosmology  Astroparticle Phys.} 10:050 (2014)

\bibitem{ber67} Bernas, R., Gradsztajn, E., Reeves, H., \& Schatzman, E. \textit{ Annals of Physics} 44:426 (1967)

\bibitem{woo90} Woosley, S.~E., Hartmann, D.~H., Hoffman, R.~D., \& Haxton, W.~C. \textit{Astrophys. J.} 356:272 (1990)

\bibitem{nak10} Nakamura, K., Yoshida, T., Shigeyama, T., \& Kajino, T. \textit{Astrophys. J.} 718:L137 (2010)

\bibitem{asp06} Asplund, M., Lambert, D.~L., Nissen, P.~E., Primas, F., \& Smith, V.~V. \textit{Astrophys. J.} 644:229 (2006)

\bibitem{jed09} Jedamzik, K., \& Pospelov, M. \textit{New J. Phys.} 11:105028 (2009)

\bibitem{rol06} Rollinde, E., Vangioni, E., \& Olive, K.~A. \textit{Astronom. J.} 651:658 (2006)

\bibitem{tat07} Tatischeff, V., \& Thibaud, J.-P. \textit{Astron. Astrophys.} 469:265 (2007)

\bibitem{lin13} Lind, K., Melendez, J., Asplund, M., Collet, R., \& Magic, Z. \textit{Astron. Astrophys.} 554:A96 (2013)

\bibitem{boe93} Boesgaard, A.~M., \& King, J.~R. \textit{Astronom. J.} 106:2309 (1993)

\bibitem{dun97} Duncan, D.~K., Primas, F., Rebull, L.~M., et al.\ 1997 \textit{Astrophys. J.} 488:338 (1997)

\bibitem{boe11} Boesgaard, A.~M., Rich, J.~A., Levesque, E.~M., \& Bowler, B.~P. \textit{Astrophys. J.} 743:140 (2011)

\bibitem{tan09} Tan, K.~F., Shi, J.~R., \& Zhao, G. \textit{Mon. Not. R. Astron. Soc.} 392:205 (2009)

\bibitem{smi09} Smiljanic, R., Pasquini, L., Bonifacio, P., et al. \textit{Astron. Astrophys.} 499:103 (2009)

\bibitem{pri10} Primas, F. \textit{IAU Symp.} 268:221 (2010)

\bibitem{pag} Pagel, B.~E.~J. \textit{Nucleosynthesis and Chemical Evolution of Galaxies}, Cambridge, UK: Cambridge University Press (2009)  

\bibitem{woo95} Woosley, S.~E., \& Weaver, T.~A. \textit{Astrophys. J. Suppl.} 101:181 (1995)

\bibitem{ram97} Ramaty, R., Kozlovsky, B., Lingenfelter, R.~E., \& Reeves, H. \textit{Astrophys. J.} 488:730 (1997)

\bibitem{dun92} Duncan, D.~K., Lambert, D.~L., \& Lemke, M. \textit{Astrophys. J.} 401:584 (1992)

\bibitem{mic13} Micali, A., Matteucci, F., \& Romano, D. \textit{Mon. Not. R. Astron. Soc.} 436:1648 (2013)

\bibitem{fie99a} Fields, B.~D., \& Olive, K.~A. \textit{Astrophys. J.} 516:797 (1999)

\bibitem{fie99b} Fields, B.~D., \& Olive, K.~A. \textit{New Astro.} 4:255 (1999)

\bibitem{cha95} Chaussidon, M., \& Robert, F. \textit{Nature}  374:337 (1995)

\bibitem{lam98} Lambert, D.~L., Sheffer, Y., Federman, S.~R., et al. \textit{Astrophys. J.} 494:614 (1998)

\bibitem{yos08} Yoshida, T., Suzuki, T., Chiba, S., et al. \textit{Astrophys. J.} 686:448 (2008) 

\bibitem{rea84} Read, S.~M., \& Viola, V.~E., Jr. \textit{Atomic Data and Nuclear Data Tables} 31:359 (1984)

\bibitem{mer01} Mercer, D.~J., Austin, S.~M., Brown, J.~A., et al. \textit{Phys. Rev.} C63:065805 (2001)

\bibitem{abi88} Abia, C., \& Canal, R. \textit{Astron. Astrophys.} 189:55 (1988)

\bibitem{par99a} Parizot, E., \& Drury, L. \textit{Astron. Astrophys.} 346:329 (1999)

\bibitem{par99b} Parizot, E., \& Drury, L. \textit{Astron. Astrophys.} 346:686 (1999)

\bibitem{tat11} Tatischeff, V., \& Kiener, J. \textit{Memorie della Societa Astronomica Italiana} 82:903 (2011)

\bibitem{tat14} Tatischeff, V., Duprat, J., \& de S{\'e}r{\'e}ville, N. \textit{Astrophys. J.} 796:124 (2014)

\bibitem{fel94} Feltzing, S., \& Gustafsson, B. \textit{Astrophys. J.} 423:68 (1994)

\bibitem{pra93b} Prantzos, N., Casse, M., \& Vangioni-Flam, E. \textit{Origin and Evolution of the Elements} p. 156 (1993b)

\bibitem{ces81} Cesarsky, C.~J., \& Bibring, J.-P., in \textit{Origin of Cosmic Rays}, IAU Symp. 94:361 (1981)

\bibitem{lin98} Lingenfelter, R.~E., Ramaty, R., \& Kozlovsky, B. \textit{Astrophys. J.}  500:L153 (1998)

\bibitem{nay96} Naya, J.~E., Barthelmy, S.~D., Bartlett, L.~M., et al. \textit{Nature} 384:44 (1996)

\bibitem{die06} Diehl, R., Halloin, H., Kretschmer, K., et al. \textit{Astron. Astrophys.} 449:1025 (2006)

\bibitem{zir10} Zirakashvili, V.~N., \& Aharonian, F.~A. \textit{Astrophys. J.}  708:965 (2010)

\bibitem{mey99} Meyer, J.-P., \& Ellison, D.~C. \textit{LiBeB Cosmic Rays, and Related X- and Gamma-Rays} 171:187 (1999)

\bibitem{ell99} Ellison, D.~C., \& Meyer, J.-P. \textit{LiBeB Cosmic Rays, and Related X- and Gamma-Rays} 171:207 (1999) 

\bibitem{cas75} Cass\'e, M., \& Soutoul, A. \textit{Astrophys. J.}  200:L75 (1975)

\bibitem{isr05} Israel, M.~H., Binns, W.~R., Cummings, A.~C., et al. \textit{Nucl. Phys.} A758:201 (2005)

\bibitem{ner16} Neronov, A., \& Meynet, G. \textit{Astron. Astrophys.} 588:A86 (2016)

\bibitem{don12} Donnelly, J., Thompson, A., O'Sullivan, D., et al. \textit{Astrophys. J.} 747:40 (2012) 

\bibitem{ale16} Alexeev, V., Bagulya, A., Chernyavsky, M., et al. \textit{Astrophys. J.} 829:120 (2016)

\bibitem{pia17} Pian, E., D'Avanzo, P., Benetti, S., et al. \textit{Nature} 551:67 (2017)

\bibitem{bin08} Binns, W.~R., Wiedenbeck, M.~E., Arnould, M., et al. \textit{New Astro. Rev.} 52:427 (2008)

\bibitem{cas82} Casse, M., \& Paul, J.~A. \textit{Astrophys. J.} 258:860 (1982)

\bibitem{pra12a} Prantzos, N. \textit{Astron. Astrophys.} 538:A80 (2012) 

\bibitem{hir05} Hirschi, R., Meynet, G., \& Maeder, A. \textit{Astron. Astrophys.} 433:1013 (2005)

\bibitem{hir06} Hirschi, R., \& et al. \textit{Rev. Modern Astron.}, 19:101 (2006)

%\bibitem{cap11} Caprioli, D. \textit{J. Cosmology Astroparticle Phys.} 5:026 (2011)

%\bibitem{gab17} Gabici, S. \textit{6th International Symposium on High Energy Gamma-Ray Astronomy} 1792:020002 (2017)

\bibitem{tim95} Timmes, F.~X., Woosley, S.~E., \& Weaver, T.~A. \textit{Astrophys. J. Suppl.} 98:617 (1995)

\bibitem{che10} Cherchneff, I. \textit{Hot and Cool: Bridging Gaps in Massive Star Evolution} 425:237 (2010)

\bibitem{sma15} Smartt, S.~J. \textit{Publications of the Astronomical Society of Australia} 32:e016 (2015)

\bibitem{par99c} Parizot, E., \& Drury, L. \textit{Astron. Astrophys.} 349:673 (1999c)

\bibitem{axf81} Axford, W.~I. \textit{Annals of the New York Academy of Sciences} 375:297 (1981)

\bibitem{pra05} Prantzos, N. in \textit{Chemical Abundances and Mixing in Stars in the Milky Way and its Satellites}, ESO Astrophysics Symposia (Springer-Verlag), 351 (2005)

\bibitem{ali02} Alib{\'e}s, A., Labay, J., \& Canal, R. \textit{Astrophys. J.} 571:326 (2002)

\bibitem{lin07a} Lingenfelter, R.~E., \& Higdon, J.~C. \textit{Astrophys. J.} 660:330 (2007)

\bibitem{deh14} De Horta, A.~Y., Sommer, E.~R., Filipovi{\'c}, M.~D., et al. \textit{Astronomical. J.} 147:162 (2014)

\bibitem{bal06} Baldi, A., Raymond, J.~C., Fabbiano, G., et al. \textit{Astrophys. J.} 636:158 (2006)

\bibitem{lin07b} Lingenfelter, R.~E., \& Higdon, J.~C. \textit{Space Science Reviews} 130, 465 (2007)

\bibitem{hig05} Higdon, J.~C., \& Lingenfelter, R.~E. \textit{Astrophys. J.} 628:738 (2005)

\bibitem{hig03} Higdon, J.~C., \& Lingenfelter, R.~E. \textit{Astrophys. J.} 590:822 (2003)

\bibitem{gar79} Garcia-Munoz, M., Simpson, J.~A., \& Wefel, J.~P. \textit{Astrophys. J.} 232:L95 (1979)

\bibitem{cas95} Cass{\'e}, M., Lehoucq, R., \& Vangioni-Flam, E. \textit{Nature}  373:318 (1995)

\bibitem{van96} Vangioni-Flam, E., Casse, M., Fields, B.~D., \& Olive, K.~A. \textit{Astrophys. J.} 468:199 (1996) 

\bibitem{van98} Vangioni-Flam, E., Ramaty, R., Olive, K.~A., \& Casse, M. \textit{Astron. Astrophys. Rev.} 337:714 (1998)

\bibitem{lem98} Lemoine, M., Vangioni-Flam, E., \& Cass{\'e}, M. \textit{Astrophys. J.} 499:735 (1998)

\bibitem{fuc06} Fuchs, B., Breitschwerdt, D., de Avillez, M.~A., Dettbarn, C., \& Flynn, C. \textit{Mon. Not. R. Astron. Soc.} 373:993 (2006)

\bibitem{ind15} Indriolo, N., Neufeld, D.~A., Gerin, M., et al. \textit{Astrophys. J.} 800:40 (2015)

%\bibitem{ind10} Indriolo, N., Blake, G.~A., Goto, M., et al. \textit{Astrophys. J.} 724:1357 (2010)

\bibitem{pad09} Padovani, M., Galli, D., \& Glassgold, A.~E. \textit{Astron. Astrophys.} 501:619 (2009)

\bibitem{ben13} Benhabiles-Mezhoud, H., Kiener, J., Tatischeff, V., \& Strong, A.~W. \textit{Astrophys. J.}  763:98 (2013)

\bibitem{ram79} Ramaty, R., Kozlovsky, B., \& Lingenfelter, R.~E. \textit{Astrophys. J.} 40:487 (1979)

\bibitem{eastrogam} De Angelis, A., Tatischeff, V., Tavani, M., et al. \textit{Exp. Astron.} 44:25 (2017)

\bibitem{hessLMC} H.E.S.S.~Collaboration, Abramowski, A., Aharonian, F., et al. \textit{Science} 347:406 (2015)

\end{thebibliography}
\end{document}